\documentclass[twocolumn]{aastex631}

\usepackage{amsmath}
\usepackage{mathtools}

\shorttitle{Mid-IR-Selected TDEs}
\shortauthors{Masterson et al.}

\begin{document}

\title{A New Population of Mid-Infrared-Selected Tidal Disruption Events: Implications for Tidal Disruption Event Rates and Host Galaxy Properties}

\author[0000-0003-4127-0739]{Megan Masterson}
\affiliation{MIT Kavli Institute for Astrophysics and Space Research, Massachusetts Institute of Technology, Cambridge, MA 02139, USA}

\author[0000-0002-8989-0542]{Kishalay De}
\altaffiliation{NASA Einstein Fellow}
\affiliation{MIT Kavli Institute for Astrophysics and Space Research, Massachusetts Institute of Technology,  Cambridge, MA 02139, USA}

\author{Christos Panagiotou}
\affiliation{MIT Kavli Institute for Astrophysics and Space Research, Massachusetts Institute of Technology,  Cambridge, MA 02139, USA}

\author[0000-0003-0172-0854]{Erin Kara}
\affiliation{MIT Kavli Institute for Astrophysics and Space Research, Massachusetts Institute of Technology,  Cambridge, MA 02139, USA}

\author[0000-0001-7090-4898]{Iair Arcavi}
\affiliation{School of Physics and Astronomy, Tel Aviv University, Tel Aviv 69978, Israel}

\author[0000-0003-2895-6218]{Anna-Christina Eilers}
\affiliation{MIT Kavli Institute for Astrophysics and Space Research, Massachusetts Institute of Technology, Cambridge, MA 02139, USA}

\author[0000-0002-7197-9004]{Danielle Frostig}
\affiliation{MIT Kavli Institute for Astrophysics and Space Research, Massachusetts Institute of Technology, Cambridge, MA 02139, USA}

\author[0000-0003-3703-5154]{Suvi Gezari}
\affiliation{Space Telescope Science Institute, 3700 San Martin Drive, Baltimore, MD 21218, USA}
\affiliation{Bloomberg Center for Physics and Astronomy, Johns Hopkins University, 3400 N. Charles St., Baltimore, MD 21218, USA}

\author{Iuliia Grotova}
\affiliation{Max-Planck-Institut f\"{u}r extraterrestrische Physik, Giessenbachstrasse 1, 85748 Garching, Germany}

\author[0000-0003-3014-8762]{Zhu Liu}
\affiliation{Max-Planck-Institut f\"{u}r extraterrestrische Physik, Giessenbachstrasse 1, 85748 Garching, Germany}

\author[0000-0002-8851-4019]{Adam Malyali}
\affiliation{Max-Planck-Institut f\"{u}r extraterrestrische Physik, Giessenbachstrasse 1, 85748 Garching, Germany}

\author[0000-0002-1125-7384]{Aaron M. Meisner}
\affiliation{NSF's National Optical-Infrared Astronomy Research Laboratory, 950 N. Cherry Ave., Tucson, AZ 85719, USA }

\author[0000-0002-0761-0130]{Andrea Merloni}
\affiliation{Max-Planck-Institut f\"{u}r extraterrestrische Physik, Giessenbachstrasse 1, 85748 Garching, Germany}

\author[0000-0001-9570-0584]{Megan Newsome}
\affiliation{Las Cumbres Observatory, 6740 Cortona Drive, Suite 102, Goleta, CA 93117-5575, USA}
\affiliation{Department of Physics, University of California, Santa Barbara, CA 93106-9530, USA}

\author[0000-0001-5990-6243]{Arne Rau}
\affiliation{Max-Planck-Institut f\"{u}r extraterrestrische Physik, Giessenbachstrasse 1, 85748 Garching, Germany}

\author[0000-0003-3769-9559]{Robert A. Simcoe}
\affiliation{MIT Kavli Institute for Astrophysics and Space Research, Massachusetts Institute of Technology, Cambridge, MA 02139, USA}

\author[0000-0002-3859-8074]{Sjoert van Velzen}
\affiliation{Leiden Observatory, Leiden University, PO Box 9513, 2300 RA Leiden, The Netherlands}

\correspondingauthor{Megan Masterson}
\email{mmasters@mit.edu}

\begin{abstract}

Most tidal disruption events (TDEs) are currently found in time-domain optical and soft X-ray surveys, both of which are prone to significant obscuration. The infrared (IR), however, is a powerful probe of dust-enshrouded environments, and hence, we recently performed a systematic search of NEOWISE mid-IR data for nearby, obscured TDEs within roughly 200 Mpc. We identified 18 TDE candidates in galactic nuclei, using difference imaging to uncover nuclear variability amongst significant host galaxy emission. These candidates were selected based on the following IR light curve properties: (1) $L_\mathrm{W2}\gtrsim10^{42}$~erg~s$^{-1}$ at peak, (2) fast rise, followed by a slow, monotonic decline, (3) no significant prior variability, and (4) no evidence for AGN activity in WISE colors. The majority of these sources showed no variable optical counterpart, suggesting that optical surveys indeed miss numerous obscured TDEs. Using narrow line ionization levels and variability arguments, we identified 6 sources as possible underlying AGN, yielding a total of 12 TDEs in our gold sample. This gold sample yields a lower limit on the IR-selected TDE rate of $2.0\pm0.3\times10^{-5}$~galaxy$^{-1}$~year$^{-1}$ ($1.3\pm0.2\times10^{-7}$~Mpc$^{-3}$~year$^{-1}$), which is comparable to optical and X-ray TDE rates. The IR-selected TDE host galaxies do not show a green valley overdensity nor a preference for quiescent, Balmer strong galaxies, which are both overrepresented in optical and X-ray TDE samples. This IR-selected sample represents a new population of dusty TDEs that have historically been missed by optical and X-ray surveys and helps alleviate tensions between observed and theoretical TDE rates and the so-called missing energy problem. 

\end{abstract}


\section{Introduction} \label{sec:intro}

Tidal disruption events \citep[TDEs;][]{Rees1988,Gezari2021}, which occur when a star winds up so close to a supermassive black hole (SMBH) that the tidal forces exceed the star's self-gravity, provide a means by which to study the properties of previously dormant SMBHs \citep[e.g. mass and spin;][]{Kesden2012,Mockler2019,Pasham2019,Wen2022} and their nuclear environments \citep{Alexander2020}. A sizeable fraction of the disrupted star becomes bound to the SMBH, leading to a short-lived episode of accretion that is revealed via emission across the electromagnetic spectrum. Historically, the first TDEs were discovered as luminous, soft X-ray flares with the ROSAT all-sky X-ray survey in the 1990s \citep[e.g.][]{Bade1996,Komossa1999a,Komossa1999b,Donley2002}, with this transient soft X-ray emission likely coming from the formation of an accretion disk. A decade later, transient UV emission, also thought to be associated with the formation of an accretion disk, was discovered with the GALEX satellite, revealing a handful of TDEs \citep{Gezari2006,Gezari2008,Gezari2009}. 

The discovery of optical emission from TDEs \citep[e.g.][]{vanVelzen2011,Gezari2012,Arcavi2014,Chornock2014,Holoien2014}, however, has led to an explosion in the TDE discovery rate. Optical emission in TDEs does not come directly from the formation of an accretion disk, but is instead thought to be associated either with the reprocessing of UV/X-ray emission from the accretion disk in surrounding optically thick material \citep{Guillochon2013,Roth2016} or with shocks from the intersection of stellar debris streams \citep{Piran2015}. With increasing cadence and depth, optical surveys like the All-Sky Automated Survey for Supernovae \citep[ASAS-SN;][]{Shappee2014}, Asteroid Terrestrial-impact Last Alert
System \citep[ATLAS;][]{Tonry2018}, and the Zwicky Transient Facility \citep[ZTF;][]{Bellm2019} now serve as the primary discovery method for TDEs, finding on the order of ten TDEs per year. In addition, SRG/eROSITA \citep{Predehl2021}, the first all-sky survey in the X-rays since ROSAT, has also been used to discover numerous new TDE candidates \citep{Sazonov2021,Malyali2021,Malyali2023a,Malyali2023b,Liu2023}. Future surveys, including the Legacy Survey of Space and Time (LSST) at the Vera Rubin Observatory \citep{Ivezic2019}, are expected to increase this discovery rate to on the order of hundreds to thousands of TDEs per year \citep[e.g.][]{Bricman2020}.

Both optical and X-ray surveys have brought TDEs into the era of demographic and population studies \citep[e.g.][]{Sazonov2021,vanVelzen2021c,Hammerstein2023,Yao2023}. One notable finding of these population studies is that the optical TDEs are preferentially found in E+A or quiescent Balmer-strong galaxies, which are a relatively rare type of galaxy characterized by no ongoing star formation, but with evidence for a recent burst of star formation less than 1 Gyr ago \citep{Arcavi2014,French2016}. Likewise, optical and X-ray TDEs are also often found in the so-called ``green valley" of the galaxy color-mass diagram \citep{Law-Smith2017,Hammerstein2021,Sazonov2021,vanVelzen2021c,Hammerstein2023,Yao2023}, which is proposed to be a transitional region between star forming and passive \citep[e.g.][]{Martin2007}, although the exact way in which galaxies transition is still debated \citep[e.g.][]{Schawinski2014}. These findings, together with a preference for centrally-concentrated TDE host galaxies \citep{Law-Smith2017,Graur2018,French2020,Hammerstein2021}, have been used to argue that TDEs occur preferentially in galaxies which have undergone a recent merger \citep[e.g.][]{Pfister2019,Hammerstein2021}. On the contrary, \cite{Roth2021} argued that dust obscuration can explain the lack of TDEs from ZTF in star-forming galaxies, suggesting that the above findings could be affected by biased selection with optical surveys. Thus, to assess the true demographics of TDEs, it is crucial to probe and select TDEs in other wavebands, like the infrared (IR), that do not miss events due to heavy dust obscuration.

Another puzzle arising from population studies of optical and X-ray selected TDEs is that the observed rate seems to be systematically lower than what is expected based on loss-cone dynamics \citep[e.g.][]{Wang2004,Stone2016}. The first observed TDE rates fell roughly an order of magnitude lower than the expected theoretical rate of 10$^{-4}$ galaxy$^{-1}$ year$^{-1}$ \citep{Donley2002,Gezari2009,vanVelzen2014}. While the latest measurements are getting closer to the theoretically expected rate, the most recent flux-limited sample from ZTF still suggests roughly a factor of 3 discrepancy, finding an average rate of $3.2_{-1.2}^{+1.8} \times 10^{-5}$ galaxy$^{-1}$ year$^{-1}$ \citep{Yao2023}. There is, however, some uncertainty on the theoretical rates, as recent works have suggested that effects such as scattering and collisions can decrease the expected rate to match that of optical and X-ray selected TDEs \citep[e.g.][]{Teboul2024}. That said, the observed rates and SMBH mass functions constructed from TDEs have important implications for black hole spin distributions \citep[e.g.][]{Kesden2012}, particularly around high-mass SMBHs as only the most rapidly spinning SMBHs above $M \gtrsim 10^8 \, M_\odot$ can disrupt a Sun-like star. Likewise, the observed rates of TDEs can help constrain the low-mass SMBH occupation fraction \citep[e.g.][]{Stone2016}, which is currently poorly constrained by other methods. Therefore, a full census of TDEs across the entire electromagnetic spectrum is vital to our ability to use TDEs as probes of SMBH demographics.

To that end, the variable IR sky is one thus-far under-utilized way to search for previously missing TDEs. IR follow-up of optically-discovered TDEs has shown the presence of numerous luminous IR flares \citep[e.g.][]{Dou2016,Jiang2016,vanVelzen2016,Newsome2023}, thought to be associated with the reprocessing of optical, UV, and X-ray emission by circumnuclear dust \citep{Lu2016,vanVelzen2021b}. The majority of optical TDEs with a dust echo have been shown to have a relatively low dust covering factor \citep[$\sim$1\%; ][]{vanVelzen2016,Jiang2021a}. However, if the dust covers a sizable fraction of the sky and our sight line to the intrinsic flare emission, then we may only see reprocessed IR emission. A population of IR-only TDEs has significant implications for TDE demographics, including affecting the observed rates and potentially explaining the lack of optical and X-ray selected TDEs in dusty, star-forming galaxies.

Several past works have attempted to find this population of obscured TDEs, using a variety of methods. Recent efforts targeting dust-obscured supernovae in galactic nuclei revealed a few TDEs \citep[e.g.][]{Mattila2018, Kool2020}, but these studies probe TDE demographics in rare environments. Other efforts, most notably the Mid-InfraRed Outbursts in Nearby Galaxies (MIRONG) sample from \cite{Jiang2021b,Wang2022b}, have focused on searching in known spectroscopic galaxies for large amplitude flares ($> 0.5$ mag) in the integrated mid-IR luminosity reported in the NEOWISE survey source catalogs. While this approach has revealed a number of potential TDE candidates, it is prone to missing events due to nuclear variability being diluted by the stellar light in integrated photometry. Additionally, the MIRONG sample is also restricted to Sloan Digital Sky Survey (SDSS) galaxies within $z < 0.35$ and is agnostic to the existence of an active galactic nucleus (AGN), thereby restricting the search area significantly and potentially introducing a significant number of AGN flares into the sample.

In this work, we aim to produce a less biased sample of IR-selected TDEs by utilizing novel difference imaging techniques for candidate selection and a less restrictive cross-matching process that covers the entire sky. This new selection process has revealed a total of 18 IR TDE candidates, including the closest TDE discovered to date, located in a star-forming galaxy at 42 Mpc \citep{Panagiotou2023}. We present the selection of TDEs from the WISE survey in Section \ref{sec:cand_sel} and provide details of our multi-wavelength follow-up efforts in Section \ref{sec:follow-up}. The details of the IR flares, energetics, and multi-wavelength evidence for dust-enshrouded accretion flows are given in Section \ref{sec:flare}, and the host galaxy properties of this sample are explored and compared to optical and X-ray samples in Section \ref{sec:hosts}. We discuss potential sample contaminants, including AGN and supernovae (SNe) in Section \ref{sec:contaminants}, and present rates of IR-selected TDEs from this sample in Section \ref{sec:rates}. Finally, we discuss our findings in Section \ref{sec:disc} and summarize them in Section \ref{sec:summary}. Throughout this work, we assume $\Omega_M = 0.3$, $\Omega_\Lambda = 0.7$, and $H_0 = 70$\,km\,s$^{-1}$\,Mpc$^{-1}$. Unless otherwise noted, quoted uncertainties represent 90\% confidence intervals. All reported magnitudes are in the AB system and have been corrected for Galactic extinction using the \cite{Schlafly2011} dust maps and the \cite{Cardelli1989} extinction curves.

\section{Selecting Infrared TDE Candidates with WISE} \label{sec:cand_sel}

The Wide-field Infrared Survey Explorer \citep[WISE;][]{Wright2010} satellite was launched in 2009 and completed its primary mission in 2010, after surveying the entire mid-IR sky once in each of its four bands. In 2013, NASA re-activated the spacecraft as the NEOWISE mission \citep{Mainzer2014}, with the goal of studying near-Earth objects (NEOs) by continuing to survey the entire IR sky every six months in only the two shortest wavelength bands. Thanks to this continued survey, numerous long-duration IR transients, like dust echoes around optical and X-ray TDEs, have been discovered \citep[e.g.][]{Dou2016,Jiang2016,vanVelzen2016,Jiang2021a,Newsome2023}. In this work, we utilize NEOWISE data to identify dust-obscured TDEs that were missed at other wavelengths. 

We carried out a systematic search of the NEOWISE survey data for transients by performing image subtraction directly on the time-resolved, coadded W1 (3.4\,$\mu$m) and W2 (4.6\,$\mu$m) data released as part of the unWISE project \citep{Lang2014,Meisner2018}. We used a customized code \citep{De2020}, which is based on the ZOGY algorithm \citep{Zackay2016} to perform the image subtraction. A visual representation of the difference imaging process can be seen in Figure 1 of \cite{Panagiotou2023}. Each epoch of unWISE stacks of NEOWISE data consist of 12 exposures acquired at the same sky position over a duration of $\approx 1$\,day for most of the sky. Each NEOWISE exposure is 7.7\,sec, amounting to a total exposure time of $\approx 92.4$\,s for most of the sky (the effective exposure is higher near the ecliptic poles). WISE images are frequently affected by scattered light (e.g. from the moon), producing time-varying artifacts that appear in the difference images. With empirical tests while creating the pipeline, we established that these artifacts dominate the source catalog at the $5-6\,\sigma$ level. Therefore, to avoid being overwhelmed by artifacts, we chose to adopt a $7\sigma$ threshold for WISE transients. Further details of this pipeline will be presented in K. De et al. (in preparation), but importantly we note that we used the full-depth unWISE coadds from 2010-2011 as the reference images for this search.

To identify nuclear transients, we cross-matched our WISE transient sample to the CLU-compiled catalog, which contains spectroscopically-confirmed galaxies within roughly\,200 Mpc compiled from numerous existing databases. For more details on the CLU-compiled catalog, we refer the reader to Section 5.3 of \cite{Cook2019}. Specifically, we required the WISE transient to occur within 2\arcsec\, of the nucleus of a CLU galaxy. The accuracy with which we can centroid a source is much better than the PSF (which is $\approx 6-7$\arcsec\, for WISE W1 and W2), and hence, we expect to be able to reach on the order of 1\arcsec\, localization for $>7\sigma$ WISE detections. Thus, we chose a cross-matching radius of 2\arcsec\, to be relatively conservative and account for potential variability of the WISE PSF across different epochs.

In total, this cross-matching yielded 2333 mid-IR nuclear transients. Many of these sources were clearly AGN that showed long-term variability or short-duration transients like supernovae that showed flares that lasted for only one or two WISE sky passes. Thus, to obtain the cleanest sample of TDEs, we implemented the following selection criteria: (1) the transient must reach a IR luminosity of $\nu L_\nu \gtrsim 10^{42}$~erg~s$^{-1}$ at 4.6~$\mu$m\footnote{Throughout the rest of this work, we refer to $\nu L_\nu$ at 4.6$\mu$m as $L_\mathrm{W2}$.}, (2) the source must be detected in both the W1 and W2 bands for at least 5 WISE epochs ($\sim$2.5 years\footnote{Although this timescale is longer than most dust echoes around optical TDEs \citep[see e.g.][]{vanVelzen2016,Jiang2021a}, this selection criterion was chosen to minimize the number of supernovae sample contaminants.}), with a light curve that shows a fast rise, followed by a slow, monotonic decay, (3) the WISE light curves should show no significant stochastic variability prior to the flare\footnote{Stochastically variable AGN would be expected to show up as both positive and negative detections in difference imaging, whereas our TDEs are only expected to show positive detections. Hence, we required that sources not show any significant negative detections.}, and (4) the pre-flare WISE colors must not exceed W1 - W2 $>$ 0.8 \citep{Stern2012} and the source must not have been detected in the ROSAT all-sky X-ray survey, both of which would indicate an underlying AGN. There are examples in the literature of candidate TDEs occurring in AGN \citep[e.g.][]{Merloni2015,Blanchard2017,Ricci2020,Ricci2021,Cannizzaro2022,Masterson2022}, but these are much rarer than AGN flares that would overwhelm our sample if included. The first two criteria primarily rule out most SNe, while the last two primarily remove AGN. In total, these selection criteria yielded 20 transients, from which we removed 2 previously-identified SNe. Beyond these cuts, we further address sample contaminants and create a gold sample of sources which we are most confident are TDEs in Section \ref{subsec:agn_contam}. 

\begin{figure*}[t!]
    \centering
    \includegraphics[width=\textwidth]{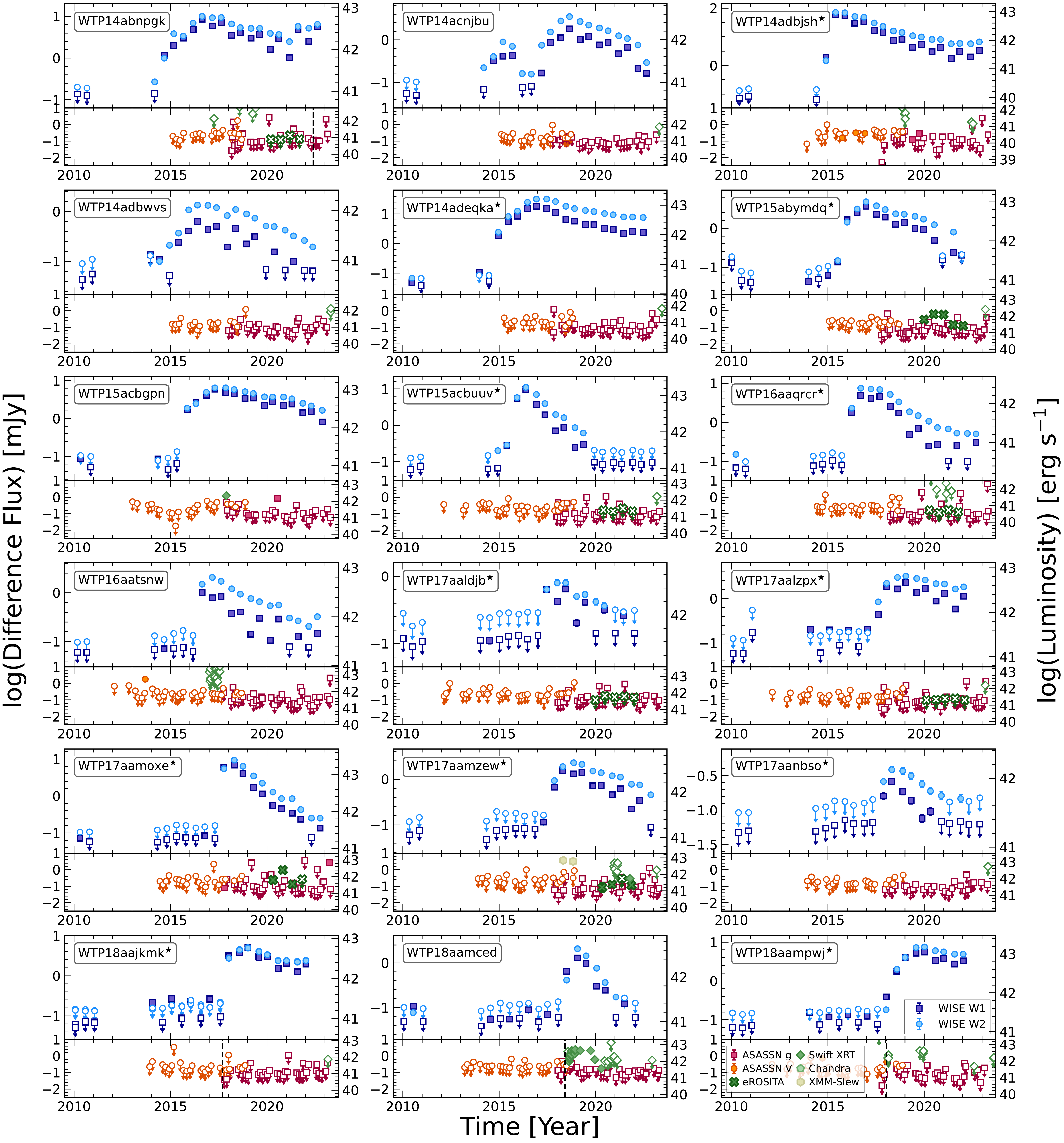}
    \caption{Multi-wavelength light curves for the 18 sources in our full sample, with gold sample sources marked with $^\bigstar$. The top panel contains the WISE data, with the purple squares and blue circles showing the WISE W1 and W2 bands, respectively. The bottom panel shows the optical and X-ray data for each source. The red squares and orange circles show the ASAS-SN $g$ and $V$ band data, respectively, binned to 60 day timescales. X-ray data is shown in green and yellow, with eROSITA shown as $\times$s, Swift XRT shown as diamonds, Chandra data shown as pentagons, and XMM-Newton Slew shown as hexagons. The X-ray data is plotted using the integrated luminosity in 0.2-2.3~keV, 0.3-10~keV, 0.5-7.0~keV, and 0.2-2~keV for eROSITA, Swift, Chandra, and XMM-Newton, respectively. For both panels, the left axis corresponds to the difference flux, whereas the right axis shows the corresponding luminosity (monochomatic at 4.6 $\mu$m for the IR, integrated for the X-ray). For all wavelengths, an empty marker with a down-pointing arrow represents an upper limit, whereas detections are shown as filled markers with error bars. Four sources show an optical transient (WTP14abnpgk, WTP18aajkmk, WTP18aamced, and WTP18aampwj), and these figures have a vertical black dashed line in the bottom panel showing the time of the optical transient in (see also Figure \ref{fig:optcounterparts}). The majority of our sources show little to no evidence for any variable optical counterparts, highlighting the importance of the mid-IR for detecting dusty TDEs.}
    \label{fig:lightcurves}
\end{figure*}

This search resulted in a total of 18 IR-selected TDE candidates, which constitute the WISE TDE sample we present in this paper. It is important to note that this is not a statistically complete sample, and there may be IR-emitting TDEs which do not exhibit the chosen selection criteria. However, this work represents the first systematic search for TDEs in the IR band with minimal sample contamination. The WISE light curves for the 18 sources are shown in Figure \ref{fig:lightcurves}, with multi-wavelength coverage shown in the bottom panels for each source. These events span a turn-on time in the IR from 2014-2018 and show little evidence for any variable optical counterparts. The brightest and closest object in our sample, WTP\,14adbjsh\footnote{Each source identified by the WISE Transient Pipeline (WTP; K. De et al. 2023, in preparation) is given a WTP name. This name consists of two letters that designate the year in which the transient was first detected in difference imaging, followed by 6 letters that indicate the order of detection (akin to the TNS naming scheme). Throughout this paper we will refer to sources by their WTP name.}, was recently published in \cite{Panagiotou2023}. Of the remaining 17 sources, 6 were detected in the MIRONG sample of IR nuclear transients \citep{Dai2020,Jiang2021b,Wang2022b}, 2 show optical flares that were detected by other groups \citep{Falco2018,Frederick2019}, and 9 are reported here for the first time. The details of our sample are given in Table \ref{tab:candidates}, and in Section \ref{sec:appendeix_indiv} of the Appendix, we describe their individual properties in detail. Follow-up for the sources was coordinated using the \texttt{fritz} astronomical data platform \citep{vanderWalt2019}. 

\begin{deluxetable*}{ccccccccc}
    \tabletypesize{\footnotesize}
	\caption{IR TDE Candidates Identified with the Wise Transient Pipeline} \label{tab:candidates}
    \tablehead{\colhead{ID} & \colhead{WTP Name} & \colhead{TNS Name\tablenotemark{$a$}} & \colhead{RA} & \colhead{Dec.} & \colhead{Redshift\tablenotemark{$b$}} & \colhead{$D_L$\tablenotemark{$c$}} & \colhead{$t_\mathrm{disrupt}$\tablenotemark{$d$}} & \colhead{First Presented} \\
    & & & (deg.) & (deg.) & & (Mpc) & {(MJD)} & }
	\startdata
	1 & WTP\,14abnpgk & AT\,2022mim & 251.477888 & -23.451942 & 0.02035 & 89.3 & 56901 & This work \\
    2 & WTP\,14acnjbu &  & 264.328943 & 19.292112 & 0.026 & 112.8 & 57823 & This work \\
    3 & WTP\,14adbjsh &  & 342.953459 & -20.608018 & 0.0106 & 41.0 & 57070 & \cite{Panagiotou2023} \\
    4 & WTP\,14adbwvs &  & 198.748944 & 51.272774 & 0.02489 & 111.1 & 56996 & \cite{Dai2020} \\
    5 & WTP\,14adeqka &  & 297.353597 & 63.509271 & 0.01895 & 80.2 & 57131 & This work \\
    6 & WTP\,15abymdq &  & 57.219702 & -55.426525 & 0.03742 & 164.3 & 57303 & This work \\
    7 & WTP\,15acbgpn &  & 166.258054 & 59.684626 & 0.03369 & 150.2 & 57310 &  \cite{Jiang2021b,Wang2022b} \\
    8 & WTP\,15acbuuv &  & 154.672586 & -13.001884 & 0.03059 & 139.5 & 57277 & This work \\
    9 & WTP\,16aaqrcr &  & 280.829573 & -62.112048 & 0.01509 & 65.0 & 57429 & This work \\
    10 & WTP\,16aatsnw &  & 254.844458 & 20.8298 & 0.04513 & 200.0 & 57495 & \cite{Jiang2021b,Wang2022b} \\
    11 & WTP\,17aaldjb &  & 197.064964 & 4.485906 & 0.04832 & 219.4 & 57765 & \cite{Jiang2021b,Wang2022b} \\
    12 & WTP\,17aalzpx &  & 50.856858 & -2.046782 & 0.03725 & 161.6 & 57972 & This work \\
    13 & WTP\,17aamoxe &  & 317.242691 & -56.475355 & 0.042 & 183.6 & 57934 & This work \\
    14 & WTP\,17aamzew &  & 146.2357 & 31.097848 & 0.03465 & 156.4 & 57944 & \cite{Jiang2021b,Wang2022b} \\
    15 & WTP\,17aanbso &  & 180.909755 & 58.986509 & 0.04692 & 210.4 & 57934 &  \cite{Jiang2021b,Wang2022b} \\
    16 & WTP\,18aajkmk &  & 40.099143 & -2.728314 & 0.0287 & 122.3  & 57991 & This work \\
    17 & WTP\,18aamced & AT\,2018dyk & 233.283417 & 44.535533 & 0.03672 & 162.9 & 58275 & \cite{Arcavi2018,Frederick2019} \\
    18 & WTP\,18aampwj & AT\,2018gn & 26.676815 & 32.508168 & 0.0375 & 161.2 & 58324 & \cite{Falco2018}; This work \\
    \enddata

    \tablenotetext{a}{For sources where an optical transient has been discovered, we list the TNS name associated with the transient in this column.}
    \tablenotetext{b}{All redshifts were taken from NED, with the exception of WTP\,14acnjbu and WTP\,17aamoxe, for which we estimate the redshift from our LDSS3 spectrum.}
    \tablenotetext{c}{Luminosity distances were computed using the redshifts (after converting to the rest-frame of the CMB) and a $\Lambda$CDM cosmology with $H_0 = 70$~km~s$^{-1}$~Mpc$^{-1}$, $\Omega_\Lambda = 0.7$, $\Omega_M = 0.3$.}
    \tablenotetext{d}{This column reports the disruption time in MJD, as estimated from our IR light curve modeling (see Section \ref{subsec:dust_dist} for more details). Throughout the rest of this work, when we report the time since disruption, we are referring to this time. For those sources with optical flares, these WISE-predicted disruption times are often later than the optical flares, as the WISE survey is shallower than most optical surveys and its cadence (6 months) is significantly longer than optical surveys ($\approx$ days). For consistency, we use these WISE-predicted times as a rough estimate of the flare time for the entire sample (i.e. irrespective of whether an optical flare is detected).}
\end{deluxetable*}

\section{Multi-Wavelength Follow-Up Observations} \label{sec:follow-up}

\subsection{Optical Spectroscopy} \label{subsec:opt_spec}

We obtained optical spectroscopic follow-up for all of the sources in our sample, with goals of identifying spectral changes and assessing the likelihood of an underlying AGN in each source. We observed two sources (WTP\,14adbjsh and WTP\,18aampwj) with the Goodman High Throughput Spectrograph \citep[GHTS;][]{Clemens2004} on the SOuthern Astrophysical Research telescope (SOAR, Program 2022B-005, 2023A007; PI: K. De) in November 2022, the first of which was presented in \cite{Panagiotou2023}. Eleven of our sources were observed with the Low Dispersion Survey Spectrograph (LDSS3) on the Magellan/Clay Telescope in early 2023 and one source was observed with the Gemini Multi-Object Spectrograph (GMOS) on Gemini-S (Program GS-2023A-Q-205; PI: K. De). The remaining four sources without post-flare optical spectroscopy were observed with the FLOYDS spectrograph on the Faulkes Telescope North (FTN) and South (FTS), which are part of the Las Cumbres Observatory Global Telescope Network \citep{Brown2013}. The exposure times, slit widths, and further details of the new observations included in this work are given in Table \ref{tab:spec}. In all observations, the long slit was centered at the position of the galaxy nucleus. All optical spectra from SOAR/GHTS, Magellan/LDSS3, and Gemini/GMOS were reduced and flux calibrated using the \texttt{pypeit} software \citep{Prochaska2020} and observations of a nearby spectroscopic standard on the same night. Optical spectra taken with FLOYDS/Las Cumbres were subsequently reduced using the \texttt{floydsspec} pipeline\footnote{\url{https://github.com/svalenti/FLOYDS_pipeline/}}, which performs flux and wavelength calibration, cosmic-ray removal, and final spectrum extraction and is described in \cite{Valenti2014}. 

\subsection{Near-Infrared Spectroscopy} \label{subsec:near-IR_spec}

For a limited number of sources in our sample, we were also able to obtain near-IR spectroscopic observations. The advantage of these observations over optical spectroscopic follow-up is that the near-IR is less affected by heavy dust obscuration and therefore allows us to potentially peer deeper into the nucleus to look for broad spectral features. We obtained near-IR spectra from both the Folded Port Infrared Echellette \citep[FIRE;][]{Simcoe2013} instrument (Echelle mode) on Magellan/Baade Telescope and SpeX instrument \citep[SXD mode;][]{Rayner2003} on the NASA Infrared Telescope Facility (IRTF), as part of program 2023A070 and 2023B046 (PI: De). In all observations, the long slit was centered at the position of the galaxy nucleus and the total exposure time was accumulated in a series of ABBA dither sequences. We also obtained observations of a telluric standard star spatially close to the science target and temporally close to the science exposures. The FIRE observations were again reduced and flux calibrated using the \texttt{pypeit} software \citep{Prochaska2020} and observations of a nearby telluric standard. The IRTF data were reduced and extracted using the \texttt{spextool} package \citep{Cushing2004}, followed by telluric correction and flux calibration using the \texttt{xtellcor} package \citep{Vacca2003}. Further details on these observations are given in Table \ref{tab:spec} in the Appendix.

\subsection{Optical Photometry} \label{subsec:optphot}

For a comparison to optical surveys, we utilize a combination of ASAS-SN, ATLAS, and ZTF data. ASAS-SN is an all-sky survey and has the longest baseline, which allows us to probe all of the IR flares in our sample and most for a significant time before the IR flare. Despite these strengths, ASAS-SN is the shallowest of the three surveys, reaching down to $V \approx$ 17 mag ($F_\nu \approx$ 0.6 mJy) in a single visit \citep{Kochanek2017}. By stacking observations, we can reach comparable fluxes to WISE ($F_\nu \approx$ 0.1 mJy), and since the majority of dust echoes around optically-selected TDEs are dimmer than the observed optical flare \citep[see e.g.][]{vanVelzen2016,Jiang2021a}, this provides sufficient constraints on the optical counterparts to these IR-selected events. We obtain forced difference photometry through the ASAS-SN SkyPatrol service \citep{Shappee2014,Kochanek2017}, and stack the resulting ASAS-SN data on 60 day timescales, using an inverse variance weighted average with iterative sigma-clipping (at the 3$\sigma$ level) for outlier rejection. For the first five sources (those starting with WTP\,14), ASAS-SN turned on roughly around the beginning of the IR flare, so we also verified that the ASAS-SN integrated photometry showed no signs of optical variability.

To compliment the ASAS-SN data and probe dimmer optical transients, we also searched ATLAS and ZTF data for corresponding optical transients. ATLAS, which reaches down to roughly 19-20 mag in the $c$- and $o$-bands, has surveyed the northern sky (dec. $\gtrsim$ -50$^\circ$) every two days since 2015, and has complimentary coverage of the southern sky starting in 2022. We obtained ATLAS data for all of our transients using the ATLAS Forced Photometry Service\footnote{\url{https://fallingstar-data.com/forcedphot/queue/}} \citep{Tonry2018,Smith2020,Shingles2021}. ZTF has surveyed the northern sky every three days since 2018, reaching down to $r \sim 20.5$ mag. Given the limited temporal coverage, ZTF is only useful for our three most recent transients, but provides the deepest coverage for these sources. We obtained ZTF data using the ZTF Forced Photometry Service \citep{Masci2019,Masci2023}. Throughout the rest of this work, we utilize the optical survey with the deepest limits and sufficient temporal coverage for each transient. Only 4 of our 18 sources show any hint of an optical transient (WTP\,14abnpgk, WTP\,18aajkmk, WTP\,18aamced, WTP\,18aampwj); we show these optical transients in more detail in Section \ref{sec:appendeix_indiv} of the Appendix. 

\subsection{X-ray Observations} \label{subsec:xray}
 
Finally, we also performed at least one X-ray observation for each of our sources after the IR flare was discovered. We utilized data from the German half of the eROSITA all-sky survey that ran from 2019-2022, and for any source not in that footprint or that did not have any archival X-ray observations, we obtained at least one Swift snapshot to estimate the late-time X-ray flux. 

\subsubsection{eROSITA} \label{subsubsec:ero}

Of the 18 sources in our sample, 8 are in the German half of the eROSITA sky. Of these 8 sources, 3 were detected during at least one eROSITA visit between 2019-2022. The eROSITA fluxes were obtained by cross-matching ($r=15\arcsec$) the IR positions with the X-ray positions from the eROSITA\_DE source catalogues for four full all-sky surveys (eRASS1-eRASS4) and a part of a fifth scan eRASS5 (eRASS1: Merloni et al. 2023; eRASS2-4: unpublished, ver.221031 and eRASS5: unpublished, ver. 221101). The fluxes and corresponding luminosity are given in 0.2-2.3 keV band, and the flux conversion factor is calculated using a power-law model with $\Gamma = 2.0$ and $N_H = 3 \times 10^{20}$ cm$^{-2}$. The eROSITA spectra were extracted using the eROSITA Science Analysis Software (eSASS) task SRCTOOL \citep[version eSASSusers\_211214;][]{Brunner2022} centred on the corresponding eRASS position. We applied SRCTOOL in AUTO mode, which determines the optimal sizes of source and background regions based on the source extension and brightness, background properties and PSF at the source position. For the non-detections, we calculated $3\sigma$ upper limits at the IR detection positions in 0.2-2.3 keV assuming the same absorbed power-law model with $\Gamma = 2.0$ and $N_H = 3 \times 10^{20}$ cm$^{-2}$ (a detailed description will be presented in I. Grotova et al. in prep.).

\subsubsection{Swift} \label{subsubsec:swiftxrt}

Many of our sources were observed with Swift \citep{Gehrels2004} at some point after their flare, either due to targeted follow-up or serendipitous observation. In addition, for any source in our sample that was not observed with other X-ray telescopes after the IR flare, we obtained one Swift Target of Opportunity observation to probe the late-time X-ray emission. With roughly 2 ks and at the distances of these transients, this allowed us to rule out any strong X-ray flux down to a limit of $F_X \approx$ few $\times 10^{-13}$ erg cm$^{-2}$ s$^{-1}$. All Swift XRT observations were reduced using the online XRT products generator\footnote{\url{https://www.swift.ac.uk/user_objects/}} \citep{Evans2007,Evans2009}. We extracted fluxes and $3\sigma$ upper limits for each Swift observation, assuming a $\Gamma = 2$ spectrum to convert from counts to flux.

\subsubsection{Chandra} \label{subsubsec:chandra}

Two sources in our sample, WTP\,15acbgpn and WTP\,16aatsnw, were observed with the Chandra X-ray Observatory during the last 12 years. We reduced these data using CIAO \citep[v4.12;][]{Fruscione2006} and CALDB (v4.9.2.1). We followed standard data reduction procedures, including running \texttt{chandra\_repro} to process the data and extracting fluxes with the \texttt{srcflux} command,  assuming a $\Gamma = 2$ spectrum. WTP\,15acbgpn was detected in 2011, more than 4 years prior to its IR flare, which suggests that this is instead an AGN, as suggested by its optical spectrum (see Section \ref{subsec:agn_contam} for more details). Despite having AGN-like ionization levels, WTP\,16aatsnw was not detected in either 2010 or 2017 with Chandra, with relatively strict X-ray flux limits of $F_{0.5-7\,\mathrm{keV}} \leq 8 \times 10^{-15}, \, 2.5 \times 10^{-14}$ erg cm$^{-2}$ s$^{-1}$ in each observation, respectively. Ultimately, both of these sources were ruled out of our gold sample due to their likely AGN nature.

\subsubsection{XMM-Newton} \label{subsubsec:xmm}

XMM-Newton performed two pointed observations of one the sources in our sample, WTP\,18aamced, which have been extensively studied in previous works \citep{Frederick2019,Huang2023}. Additionally, WTP\,17aamzew was detected twice in 2018 during the in the XMM Slew Survey \citep{Saxton2008}. We obtained the XMM-Slew fluxes from the High-Energy Light-Curve Generator\footnote{\url{http://xmmuls.esac.esa.int/hiligt/}} \citep{Konig2022,Saxton2022}, assuming a $\Gamma = 2$ power-law to convert from count rate to fluxes in the 0.2-2~keV band. The source was not detected in the 2-12~keV band, but the exposures were extremely short ($\sim$10s).

\section{Potential Sample Contaminants} \label{sec:contaminants}

To further remove potential sample contaminants and establish the cleanest possible sample of IR-selected TDEs, we break the sample into two, a gold sample, which contains the sources that we are most confident are TDEs, and a silver sample, which contains sources for which an AGN nature cannot be ruled out. In total, we find 6 sources with potential AGN contamination, which we place in the silver sample, and 12 sources which have a high likelihood of being TDEs and thereby constitute the gold sample. The designation for each source is given in Table \ref{tab:sampledesignation}, and the details of ruling out non-TDE contamination in our gold sample are given below. 

\subsection{Contamination from AGN} \label{subsec:agn_contam}

With ongoing accretion and often large amounts of dust, AGN are a natural contaminant in our sample. While AGN and TDEs are not mutually exclusive phenomena \citep[see e.g.][]{Merloni2015,Blanchard2017,Ricci2020,Ricci2021,Cannizzaro2022,Masterson2022}, AGN also often show flares that are not associated with TDEs and would therefore overwhelm our sample if we did not remove them. Hence, to produce the purest sample of TDEs, we utilize optical spectroscopy to identify and remove AGN contaminants. Spectra from before WISE flares are most useful for this exercise, as they preclude the TDEs from producing the accretion-driven emission lines. Seven sources in our sample have SDSS spectra that were taken prior to the WISE flares. For the remaining sources in our sample, we utilize our follow-up spectra from Magellan, Las Cumbres, SOAR, or Gemini to rule out AGN contribution. However, in these sources, the presence of AGN-like emission lines is difficult to disentangle from transient emission lines associated with the TDE. 

\begin{figure*}[t!]
    \centering
    \includegraphics[width=\textwidth]{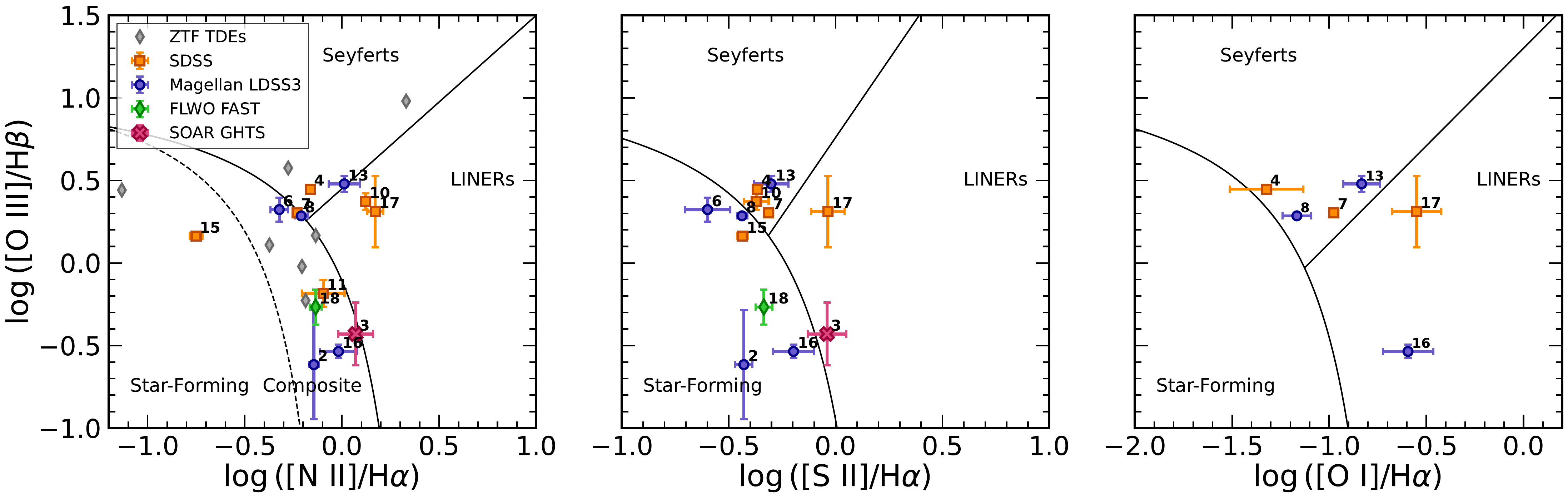}
    \caption{Various BPT diagnostic plots for the sources in our sample which have detectable emission lines. The orange square points show the archival SDSS data from the early 2000s (i.e. before the IR flares), the purple circles show the recent observations with the Magellan LDSS3 instrument in 2023, the green diamond shows an archival observation of WTP\,18aampwj from the F.L. Whipple Observatory (FLWO) FAST instrument in 2018, and the red $\times$ shows an observation of WTP\,14adbjsh from the SOAR Goodman High Throughput Spectrograph (GHTS) from 2023. We stress that all of the observations except SDSS are taken after the IR flare and may have contamination from TDE-driven ionization. The corresponding source ID number for each source (given in Table \ref{tab:candidates}) is shown to the top right of each point. Error bars represent 1$\sigma$ uncertainty. Each panel shows a different diagnostic line ratio, with the boundaries between star-forming, composite, Seyferts, and LINERs shown as either solid or dashed lines. For all sources with an archival SDSS spectrum, we only show the line ratios for the SDSS spectrum to alleviate contamination from the recent flare. The grey diamonds in the left-most panel shows the line ratio diagnostics for the 7 of 19 ZTF TDEs from \cite{Hammerstein2021} that showed significant line emission. These optically-selected TDEs show similar ionization levels compared to the WISE TDEs presented in this work.}
    \label{fig:bpt}
\end{figure*}

To accurately measure the line ratios, we first removed the stellar continuum by fitting each spectrum with penalized pixel fitting routine \citep[\texttt{pPXF};][]{Cappellari2004,Cappellari2017,Cappellari2023} using the MILES stellar library \citep{Falcon-Barroso2011}. We mask the key emission line regions, including the H$\alpha$, H$\beta$, [O III], [O I], [N II], and [S II] regions, when fitting with \texttt{pPXF}. Using the residuals of the \texttt{pPXF} fit, we fit each emission line of interest with a Gaussian model to estimate the line flux and velocity dispersions. Uncertainties on each fit parameter were determined by adding simulated noise to the spectrum, re-fitting, and repeating this process 1000 times. The noise is scaled to the root-mean-square of the residuals from \texttt{pPXF} fitting. In the majority of sources, we only see narrow emission lines, which are not resolvable with the resolution of the Magellan LDSS3 instrument. The relative line intensity ratios of these narrow lines are used to place the sources on various BPT diagrams \citep{Baldwin1981}, using three different line ratios with boundaries between star-forming, Seyfert, LINER, and composite ionization sources taken from large samples of SDSS galaxies \citep[e.g.][]{Kewley2001,Kauffmann2003,Kewley2006,Schawinski2007}. Figure \ref{fig:bpt} shows the resulting BPT diagrams, with the sources labeled based on the IDs given in Table \ref{tab:candidates}. Note that only the sources that showed strong emission lines are shown in each panel. We also show a comparison to optically-selected TDEs from ZTF \citep{Hammerstein2021}; the 7 sources with significant emission lines are consistent show a similar distribution in the BPT diagrams to the IR-selected sample. Likewise, 8 of the 13 X-ray selected TDEs from eROSITA \citep{Sazonov2021} show significant emission lines, and at least 2 of these sources are also in the LINER/Seyfert part of BPT diagrams.

We find that many of our sources do not show Seyfert or LINER ionization levels, indicating that the likelihood of previous AGN activity is low. Only 5 sources show any significant Seyfert or LINER ionization -- WTP\,14adbwvs, WTP\,15acbgpn, WTP\,16aatsnw, WTP\,18aamced, WTP\,17aamoxe. The first four of these sources were classified as AGN in SDSS with spectra taken before the IR flares, and we therefore place these sources in our silver sample, as we are less confident in their pure TDE nature. On the other hand, the spectrum of WTP\,17aamoxe used for this analysis was taken after the IR flare and hence may be contaminated by ionization from the underlying TDE. An archival spectrum from before the flare shows no AGN-driven lines, suggesting that the Seyfert-like nature of WTP\,17aamoxe may be a result of the TDE. It is important to note though that the archival spectrum was taken with larger fibers, and thus, nuclear emission lines may be diluted by host galaxy light (see Section \ref{subsec:transient_lines} for more details). However, as there are no other indicators of AGN activity in WTP\,17aamoxe, we place this source in our gold sample.

\begin{figure}[t!]
    \centering
    \includegraphics[width=0.48\textwidth]{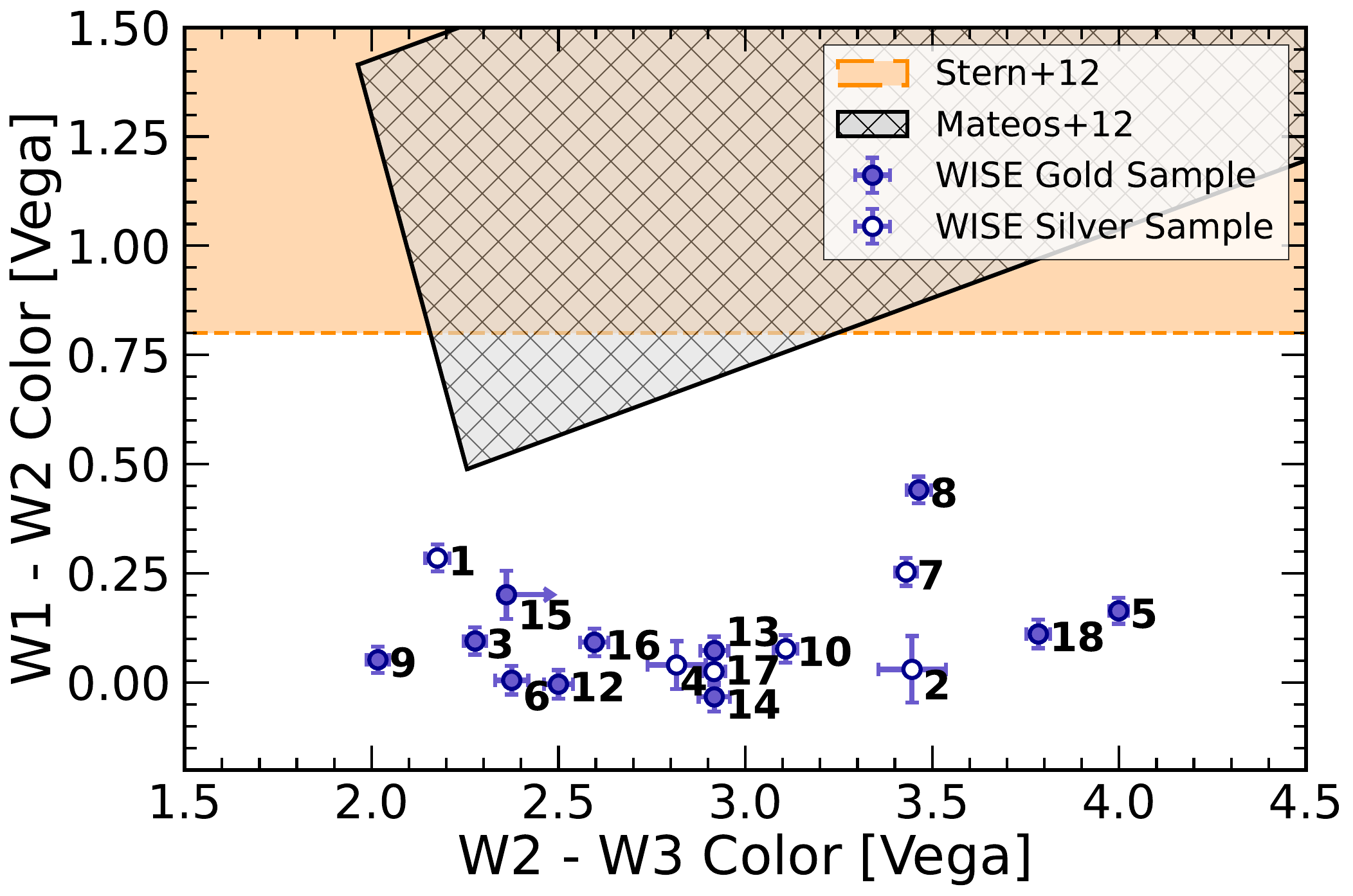}
    \caption{WISE W1-W2 color versus WISE W2-W3 color of the host galaxies, taken from the WISE catalog (i.e. from before the WISE flares). Sources in our gold sample are shown with filled markers, while sources in our silver sample are shown with empty markers. All sources are labeled to the lower right with the ID number for each source from Table \ref{tab:candidates}. The orange filled area corresponds to the AGN classification with W1-W2 $>$ 0.8 from \cite{Stern2012}, and the grey hashed region corresponds to the AGN classification from \cite{Mateos2012} using both the W1-W2 and W2-W3 colors. None of the WISE TDE candidates in our sample show evidence for AGN-like WISE colors.}
    \label{fig:wisecolors}
\end{figure}

In addition to optical spectroscopic classification, IR colors are a useful indicator of AGN activity, as they can naturally distinguish between a stellar blackbody spectrum that peaks around 1.6 $\mu$m and an AGN dust spectrum that is a power law in the near-IR \citep{Stern2005}. Having a separate indicator of AGN activity is useful, as recently there have been ``optically quiescent quasars" found, which show no narrow emission lines indicative of AGN behavior, but which show AGN WISE colors \citep[potentially due to full covering obscuring material; e.g.][]{Greenwell2021}. As is shown in Figure \ref{fig:wisecolors}, all of the sources in our sample sit far from various different WISE color cuts for AGN, including W1-W2 $>$ 0.8 from \cite{Stern2012} and a combination of W1-W2 and W2-W3 colors from \cite{Mateos2012}. The fact that most of our sample sits far from these boundaries indicates that there is not significant concern for strong AGN contamination in the majority of the sample.

Lastly, we exclude two more sources, WTP\,14abnpgk and WTP\,14acnjbu, from our gold sample, as their WISE light curves show more activity than a single flare. In particular, WTP\,14abnpgk appears to be rising again in the latest WISE data, and is actually a recent ZTF nuclear transient\footnote{\url{https://www.wis-tns.org/object/2022mim}} with broad redshifted H$\alpha$ and H$\beta$ emission (see Section \ref{sec:appendeix_indiv} in the Appendix for more details on this source). Similarly, WTP\,14acnjbu shows an initial small flare before the main flare. Such activity is suggestive of an underlying AGN, and thus, we conservatively place these two sources in our silver sample, despite the fact that their optical spectra and WISE colors do not indicate AGN activity. Additionally, the host galaxy properties of IR transients have recently been used to discriminate AGN from TDEs in the MIRONG sample \citep{Dodd2023}, although our sample suggests that IR-selected TDEs have a more diverse range of host galaxy properties than previously seen (see Section \ref{sec:hosts} and Section \ref{subsec:disc_hosts} for more details).

\subsection{Contamination from SNe} \label{subsec:sne_contam}

As some SNe also show late-time IR excesses associated either with a dust echo or newly formed dust \citep[e.g.][]{Fox2010,Meikle2011,Szalai2011,Helou2013,Ergon2015}, they are another potential contaminant to our sample. However, there are a number of reasons that SNe are not a significant contributor to our population. First, the majority of the mid-IR-detected SNe are much dimmer than what we find in our sample \citep[e.g][]{Fox2011,Tinyanont2016}, and many SNe detected in the mid-IR emit for less than three years in the MIR \citep[especially Type Ia SNe; e.g][]{Tinyanont2016}. In the MIRONG sample, which was agnostic to event type, \cite{Jiang2021b} found that the four SNe in their sample all had peak mid-IR luminosities of $L_\mathrm{W2} < 10^{42}$~erg~s$^{-1}$, which would not have been included in our sample. Additionally, very few SNe can produce the large amounts of energy radiated by the events in our sample ($E_\mathrm{rad,\, bb} \gtrsim 10^{51}$ erg for most events; see Table \ref{tab:sampledesignation} and Section \ref{subsec:ir_energetics} for more details). This lower bound on the radiated energy exceeds the standard values expected for typical SNe \citep[e.g.][]{Sukhbold2016}. More energy can be radiated in the sub-class of superluminous SNe \citep[SLSNe; for a recent review, see][]{Gal-Yam2019}, but a recent study of SLSNe in the MIR suggests that these sources are still dimmer in the MIR than accretion driven flares \citep[$L_{\mathrm{MIR}} \lesssim$ few $\times 10^{42}$ erg s$^{-1}$;][]{Sun2022}. Likewise, these events are relatively rare compared to TDEs \citep[e.g][]{Quimby2013,Frohmaier2021} and they occur primarily in low mass and low metallicity galaxies \citep[e.g.][]{Leloudas2015,Perley2016}. None of our TDE candidates reside in the type of galaxies expected to host SLSNe (see Section \ref{subsec:seds} and the left panel of Figure \ref{fig:color_mass}). Hence, we believe that there are no SLSNe contaminating our sample.

\subsection{A Clean Sample of TDEs} \label{subsec:clean_sample}

We have taken care to exclude AGN and SNe from our gold sample of TDEs, through a combination of selection criteria, optical spectroscopic analysis, WISE colors, and WISE evolution. Based on energy, rate, and host galaxy arguments, we find that there are no clear SNe in our sample. We find a few sources which have AGN-like narrow emission line ionization states before the flare, but we note that none of these sources have WISE colors indicative of AGN and the majority sit very close to the composite region parameter space. After removing these potential weak AGN, we have a total of 12 remaining TDEs in our gold sample, which represents the purest sample of IR-selected TDEs to date. All but one of these gold sample TDEs show no evidence for any optical counterpart, indicating that this sample represents a previously missed population of dust-enshrouded TDEs. Table \ref{tab:sampledesignation} gives the gold/silver designation for the entire sample, along with highlighting important criteria that went into this designation. Throughout the rest of this work, we will show both the silver and gold sample, but designate gold sources with filled markers (or mark them with $^\bigstar$) and silver sources with open markers. To reduce bias from potential AGN contamination, we only utilize the gold sample when discussing collective sample properties.

\begin{deluxetable*}{c c c c c c c c}
	\caption{Sample Designation for WISE IR TDE Candidates} \label{tab:sampledesignation}
    \tablehead{\colhead{ID} & \colhead{WTP Name} & \colhead{AGN/LINER from BPT?} & \colhead{X-rays} & \colhead{Transient Lines} & \colhead{$E_\mathrm{rad,\, bb}$ ($10^{51}$ erg)} & \colhead{Sample Designation}}
	\startdata
	1 & WTP\,14abnpgk & $\times$ & $\times$ & $\times$ & 4.3 & Silver \\
    2 & WTP\,14acnjbu & $\times$ & $\times$ & $\times$ & 1.7 & Silver \\
    3 & WTP\,14adbjsh & $\times$ & $\times$ & $\times$ & 3.7 & Gold \\
    4 & WTP\,14adbwvs & $\checkmark$ & $\times$ & $\times$ & 0.8 & Silver \\
    5 & WTP\,14adeqka & $\times$ & $\times$ & $\times$ & 9.4 & Gold \\
    6 & WTP\,15abymdq & $\times$ & $\checkmark$ & $\times$ & 4.2 & Gold \\
    7 & WTP\,15acbgpn & $\checkmark$ & $\checkmark$ & $\times$ & 7.7 & Silver \\
    8 & WTP\,15acbuuv & $\times$ & $\times$ & $\times$ & 4.0 & Gold \\
    9 & WTP\,16aaqrcr & $\times$ & $\times$ & $\times$ & 0.9 & Gold \\
    10 & WTP\,16aatsnw & $\checkmark$ & $\times$ & $\times$ & 2.8 & Silver \\
    11 & WTP\,17aaldjb & $\times$ & $\times$ & $\times$ & 1.3 & Gold \\
    12 & WTP\,17aalzpx & $\times$ & $\times$ & $\times$ & 3.2 & Gold \\
    13 & WTP\,17aamoxe & $\checkmark$ & $\checkmark$ & $\checkmark$ & 6.4 & Gold\tablenotemark{$\dagger$} \\
    14 & WTP\,17aamzew & $\times$ & $\checkmark$ & $\checkmark$ & 1.9 & Gold \\
    15 & WTP\,17aanbso & $\times$ & $\times$ & $\times$ & 0.4 & Gold \\
    16 & WTP\,18aajkmk & $\times$ & $\times$ & $\checkmark$ & 2.9 & Gold \\
    17 & WTP\,18aamced & $\checkmark$ & $\checkmark$ & $\times$ & 1.9 & Silver \\
    18 & WTP\,18aampwj & $\times$ & $\times$ & $\checkmark$ & 7.0 & Gold \\
    \enddata

    \tablenotetext{\dagger}{Despite the AGN-like ionization levels in its optical spectrum, we keep WTP\,17aamoxe as a gold TDE candidate because its archival optical spectrum shows no such lines, so the ionization of these lines could be due to a recent TDE flare. See \ref{subsec:transient_lines} and Figure \ref{fig:wtp17s} for more details.}
\end{deluxetable*}

\section{Flare Properties \& Evidence for Dust-Enshrouded Accretion} \label{sec:flare}

\subsection{IR Flare Evolution \& Energetics} \label{subsec:ir_energetics}

In Figure \ref{fig:lightcurves}, we show each IR flare from WISE, with the corresponding multi-wavelength coverage plotted in the bottom panels. As per our selection criteria, these TDE candidates show a sharp rise followed by a smooth, slow decline back to the baseline flux. They span over an order of magnitude in peak IR luminosity ($L_\mathrm{W2}$ = 10$^{42}$--$3 \times 10^{43}$~erg~s$^{-1}$), and most are still IR bright in the latest WISE epoch (second half of 2022), despite it being more than 5 years since disruption for all sources. 

The exact energetics of the flares depend on the shape of the dust SED, which depends on the dust composition, temperature, and size of the grains. Given that we only have access to two photometric bands with WISE, we have very limited knowledge of the IR spectral shape and hence, cannot disentangle different dust emission models. Therefore, we fit the WISE data with both a simple, single temperature blackbody and a modified blackbody, or ``gray body" \citep[e.g.][]{Draine1984}, with $B_\nu'(T) = \nu^{1.8} B_\nu(T)$, as in \cite{vanVelzen2016} for graphite dust grains with a grain size of $a = 0.1 \mu$m. From these fits we derive both the temperature and integrated IR luminosity,  which we can then use to infer the emitting radius of the dust (see Figure \ref{fig:dust_rad}). Following \cite{Panagiotou2023}, we adopt a 10\% systematic uncertainty on the WISE fluxes to account for the changing PSF of the telescope across different visits. To assess the uncertainty on the fit parameters, we resample the WISE data points from within their respective uncertainties and fit these new data points, repeating this process 1000 times to create a distribution of best-fit temperatures and normalizations that are consistent with the data. The majority of our sources show a very similar temperature evolution; the initial part of the flare is relatively hot ($T \approx 800-1500$ K for the blackbody model, with the modified blackbody giving temperatures that are on average 300 K lower), followed by cooling to lower temperatures ($T \approx 500-800$ K for the blackbody). This behavior can also be seen qualitatively by the increase in the W2 to W1 flux ratio during the evolution of the flares in Figure \ref{fig:lightcurves}. 

With the best-fit blackbody, we estimate the cumulative amount of energy released in the IR by each flare. The flares emit from $E_\mathrm{rad,\, bb} \approx 4 \times 10^{50}$--$9 \times 10^{51}$ erg, with no correlation between the time since the flare and the emitted energy, indicating that we are not missing a significant fraction of energy from the flares that began later. The value of $E_\mathrm{rad,\, bb}$ from the blackbody fit for each flare is reported in Table \ref{tab:sampledesignation}. However, it is important to note that these estimates are lower bounds on the total radiated energy because there may be energy emitted in other bands that is not absorbed and reprocessed by the dust (e.g. hard X-rays) and the dust may not cover the entire sky and thereby only re-emit a portion of the intrinsic optical/UV/soft X-ray flare. The covering factor is usually estimated by measuring the ratio of the intrinsic flare emission to the dust reprocessing flare, but we do not see the intrinsic flare in many of these events. However, the sharpness of the light curves imply relatively small radii, comparable to the radii from blackbody models, thereby suggesting relatively high covering factors, which we address further in the following subsection.

We find that the majority of the TDE candidates in our sample show little observable transient emission outside of the mid-IR flare identified by WISE. In particular, only three sources were detected by optical surveys -- WTP\,14abnpgk (identified as a nuclear transient by ZTF during its secondary rise in 2022), WTP\,18aampwj \citep[identified by ASAS-SN and originally classified as a SN;][]{Falco2018}, and WTP\,18aamced \citep[identified as a new changing-look LINER by ZTF;][]{Frederick2019}. Upon investigation of the ATLAS data for all 18 transients, we also identified a possible optical transient in WTP\,18aajkmk that is temporally coincident with the WISE flare. The transient is not seen in ZTF, but this could be due to timing relative to when the ZTF reference images were taken. Further details of the optical counterparts and a detailed description of each source are given in Section \ref{sec:appendeix_indiv} of the Appendix (see also Figure \ref{fig:optcounterparts}). The lack of optical counterparts in the majority of the events in our sample and the relatively dim nature of the detected optical transients relative to the IR peak suggests that the optical flux is highly attenuated in these systems.

\subsection{Dust Distributions} \label{subsec:dust_dist}

The WISE light curves encode information about the location of the emitting dust. Specifically, the rise time is set by the distance from the central irradiating source, as more distant dust leads to a longer rise time set by the extended light travel time to the far side of the dust distribution. We can therefore estimate the location of the dust with respect to the central irradiating source by modeling the WISE light curves with a simple spherical shell model for the dust, the details of which are given in \cite{vanVelzen2016}. In short, the IR light curve can be written as the convolution between a response function of the dust, $\Psi$, and the intrinsic light curve, $L_\mathrm{int}$, which is given by
\begin{equation}
    L_\mathrm{IR} = \int d\tau \, \Psi(\tau) L_\mathrm{int}(t-\tau),
\end{equation}
where $\tau = (R/c)(1-\cos\theta)$ is the delay as a function of the polar angle ($\theta$). For a simple spherical shell model for the dust, the response function is a top-hat in $\tau$, uniform from $\tau = 0$ to $\tau = 2R/c$ and 0 elsewhere. We also need to assume an underlying optical/UV flare to model the WISE light curves. As the majority of our flares have no observable optical/UV flare, we assume that the underlying flare follows a power law decline, given by
\begin{equation}
    L_\mathrm{int} = L_0 \left(\frac{t - t_p + t_0}{t_0}\right)^{-5/3},
\end{equation}
as is commonly used in modeling of the optical flares of TDEs \citep[e.g.][]{vanVelzen2021c,Hammerstein2023,Yao2023}. As the majority of the TDEs in our sample do not show a strong optical flare, we set the constant of normalization, $L_0$, to be 1 and fit for the amplitude of the resulting IR flare. This yields a model with four free parameters -- the radius of the dust shell, amplitude of the IR flare, peak time, $t_p$, and the fall-back time, $t_0$. 

\begin{figure*}[t!]
    \centering
    \includegraphics[width=\textwidth]{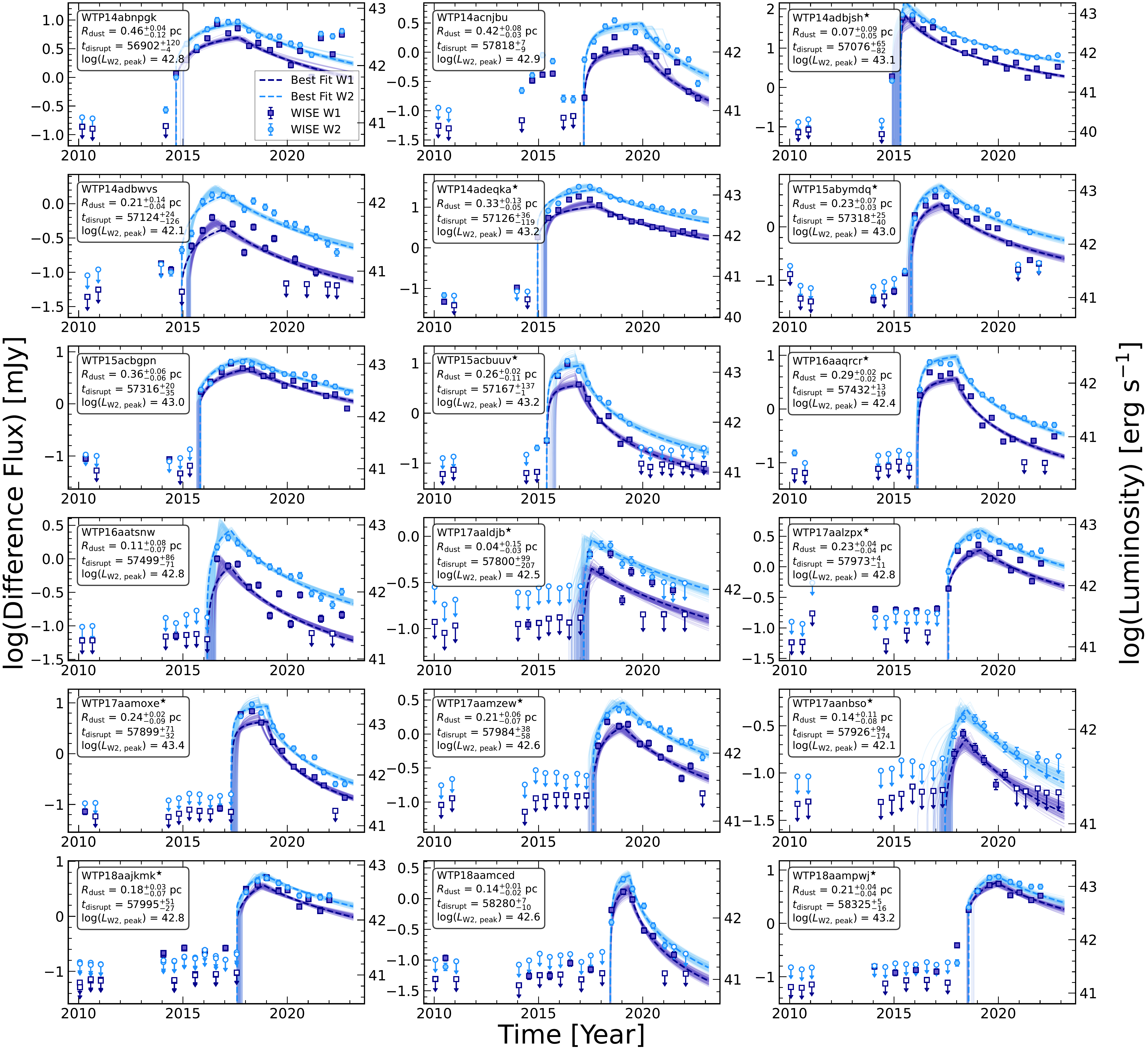}
    \caption{WISE W1 and W2 light curves modeled with a spherical shell of dust. The WISE W1 and W2 data are shown with purple squares and blue circles, respectively, with an empty marker with a down-pointing arrow represents an upper limit. For each band, the best-fitting spherical dust shell model is shown with a dashed line, while the lighter shaded lines are a random sample of 100 MCMC chain steps. The inset text gives the best fitting radius of the dust shell, disruption time, and peak W2 luminosity (in units of~erg~s$^{-1}$). The left axis corresponds to the IR difference flux, whereas the right axis shows the monochomatic luminosity at 4.6 $\mu$m ($L_\mathrm{W2}$). Sources in the gold sample are marked with $^\bigstar$ A spherical shell dust model can accurately reproduce these IR TDE flares.}
    \label{fig:dustmodel}
\end{figure*}

We fit both the W1 and W2 light curves simultaneously with this spherical shell model, with all parameters tied across the two bands. As this simple model cannot account for changing temperature, we fix the ratio between the W1 and W2 fluxes to that from a blackbody with the temperature fixed at the average across the entire flare. This provides a sufficient fit to the two bands, as there is only mild temperature evolution in these flares. Full radiative transfer is needed to account for temperature evolution, which is beyond this scope of this work, but will be explored in the future. We fit the model using the \texttt{emcee} package \citep{Foreman-Mackey2013} to implement Markov Chain Monte Carlo (MCMC) fitting routines with the likelihood in Equation (21) of \cite{vanVelzen2016} and uniform priors across all parameters. The chains contained 32 walkers and were initialized based on a maximum likelihood fit. For each walker, we burned the first 500 chain steps and then ran the chain for 5000 more steps, enough to have well-sampled and stable chains. The majority of sources showed relatively well-behaved chains, but some sources exhibited mild degeneracy between the radius and amplitude parameters, likely arising from the sparsely-sampled WISE light curves. For the purposes of this work, however, this does not pose a significant issue, as we capture the mild degeneracies in our parameter uncertainties. We show the best-fitting model, along with the model from a random sample of 100 MCMC steps, overplotted on the WISE data in Figure \ref{fig:dustmodel}.

\begin{figure*}[t!]
    \centering
    \includegraphics[width=\textwidth]{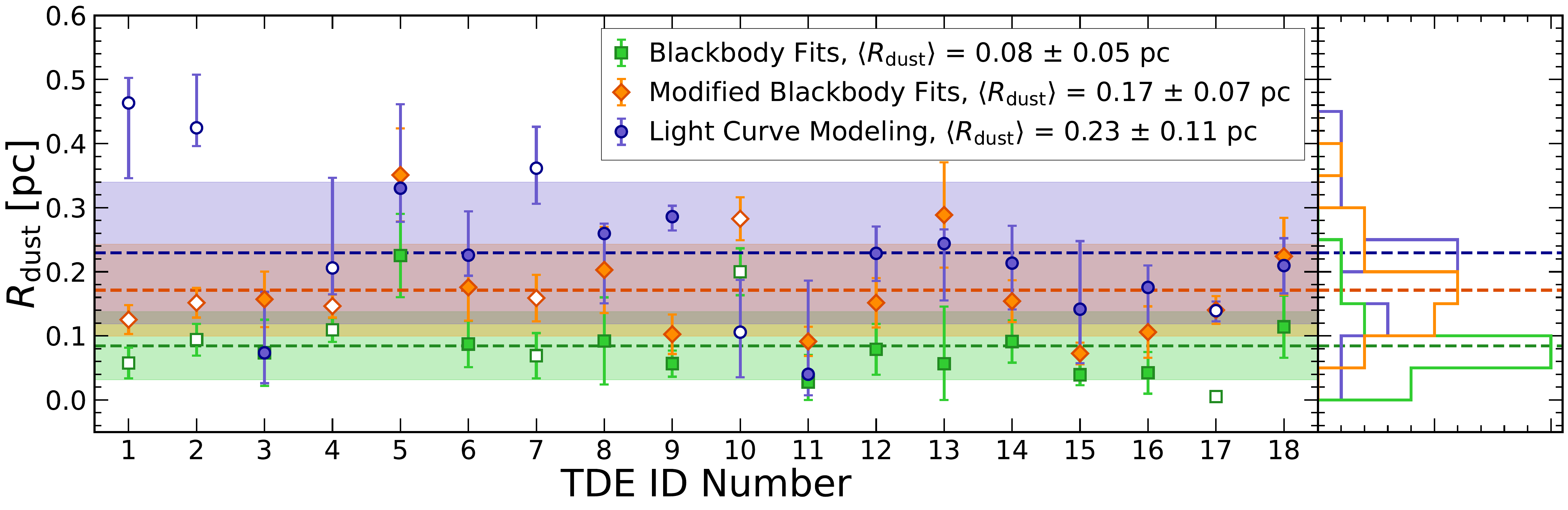}
    \caption{Different estimates of the dust radius ($R_\mathrm{dust}$) for each TDE in our sample. The purple shows the dust radius computed from light curve modeling with a spherical dust shell, the green shows the dust radius computed from the blackbody model, and the orange shows the dust radius computed from a modified blackbody. Sources in our gold sample are shown with filled markers, while sources in our silver sample are shown with empty markers. The left panel of the figure gives the individual dust radii for each TDE candidate, and the right panel shows the histogram for each estimator. The dust distance is consistently close to the blackbody/modified blackbody distance estimates, which indicates that the covering factor of the dust is close to, but slightly less than 1.}
    \label{fig:dust_rad}
\end{figure*}

The dust radii span from 0.05 to 0.46 pc with an average of $\langle R_\mathrm{dust} \rangle = 0.23 \pm 0.11$ pc, which corresponds to a range of light travel times between 60 and 550 days. It is important to note that by selecting sources whose IR light curves showed fast rise, slow decline behavior, we implicitly restricted our sample to sources where the dust was located relatively close to the nucleus as the IR rise time is set by the distance to the dust. Figure \ref{fig:dust_rad} shows the derived radii for each TDE candidate from our light curve modeling, compared with the radius estimated from both a standard and modified blackbody (i.e. $R_\mathrm{dust} = (L / 4 \pi \sigma T^4)^{1/2}$) at the peak of the IR outburst. The radius estimates from the blackbody fits are systematically smaller than the estimates based on the light curve modeling. This is expected, as a blackbody is the most efficient radiation mechanism and thereby provides an absolute lower limit on the radius. However, the radius estimates from the modified blackbody fits are between the light curve and blackbody estimates, and these estimates are often close to or within uncertainty of the light curve modeling estimates. This general agreement between the radii suggests that the covering factor is high ($f_c \gtrsim 10\%$) in these systems since the observed blackbody and modified blackbody luminosities are dependent on the covering factor. This is much higher than what is typically seen in optically-selected TDEs, where the covering factor, which can be directly measured as the IR to optical flux at peak, is on the order of 1\% \citep{Jiang2021a}. A comparably high covering factor ($f_c \sim 20\%$) has also been seen in the recent discovery of a bright MIR flare ($L_\mathrm{IR} \approx 2 \times 10^{43}$~erg~s$^{-1}$) around an optical and X-ray emitting TDE \citep{Wang2022a}. We caution that an exact measurement of the covering factor is not possible in these IR-selected TDEs, given that the optical flare is not observed in most of these systems. Estimating the covering factor from the IR light curve alone requires significant assumptions about the dust temperature, size, and composition that can lead to over an order of magnitude uncertainty. However, this analysis corroborates the idea that this sample of IR-only emitting TDEs are highly dust obscured TDEs.

\subsection{Transient Emission Lines in the Optical and Near-IR} \label{subsec:transient_lines}

\begin{figure*}[t!]
    \centering
    \includegraphics[width=\textwidth]{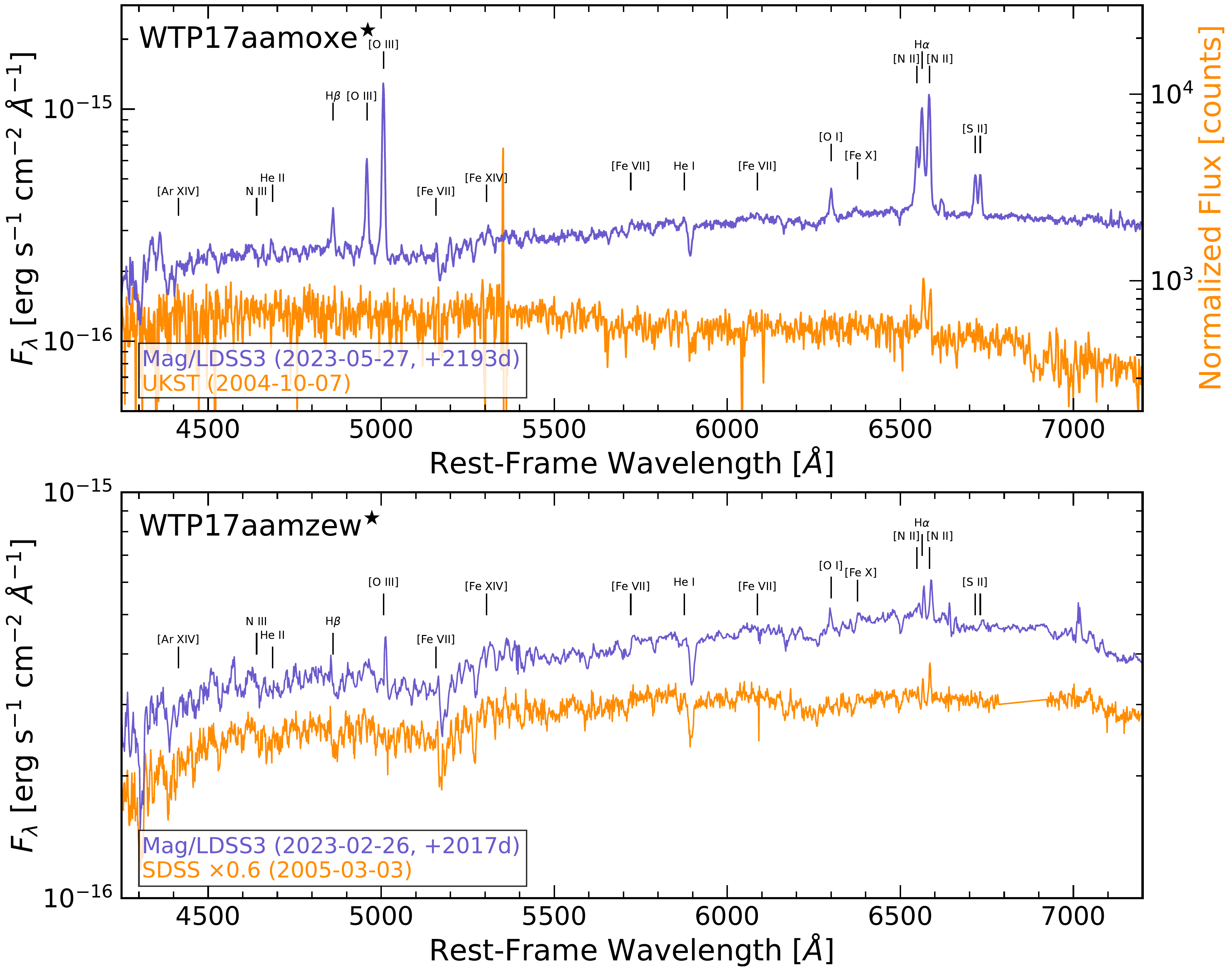}
    \caption{Evolution of transient narrow forbidden emission lines in the optical spectra of WTP\,17aamoxe (top) and WTP\,17aamzew (bottom), both of which are in our gold sample. For both sources, the purple shows recent Magellan spectra taken in 2023, both of which show prominent accretion-driven lines (e.g. [O III], [O I]). For WTP\,17aamoxe, the orange shows an archival spectrum taken on the UKST in 2004 as part of the 6dF galaxy redshift survey. The spectrum is not flux calibrated, so the flux is given in counts on the right axis and offset to highlight the difference between the two spectra. However, there is no indication of underlying accretion (i.e. no [O III], [O I], or [S II] lines) in this spectrum. This could be due to the WISE-detected TDE, a turn-on AGN scenario, although it is important to note that the UKST spectrum was taken with large (6.7\arcsec) fibers that contain significant host galaxy contributions. For WTP\,17aamzew, the orange shows the archival SDSS spectrum from 2005, with an offset to highlight the change between the two spectra. The archival SDSS spectrum has no indication of underlying accretion (i.e. no [O III], [O I], or [S II] lines), while the recent LDSS spectrum taken roughly 6 years after the initial disruption shows significant narrow [O III] and [O I] emission, which we speculate could arise from the ionization of nearby gas.}
    \label{fig:wtp17s}
\end{figure*}

With our follow-up spectroscopy program, we identified a few sources that showed evidence of spectral changes on timescales of months to years. Two sources, WTP\,17aamzew and WTP\,17aamoxe, showed the potential appearance of forbidden narrow lines commonly associated with accretion, which is highlighted in Figure \ref{fig:wtp17s}. The SDSS spectrum from 2005 of WTP\,17aamzew did not show significant emission lines from [O III], [O I], or [S II]. However, our 2023 spectrum showed that both narrow [O III] and [O I] emission lines has appeared at $>$5$\sigma$ significance and with strengths suggesting LINER-like ionization. The flux has increased from $F_\mathrm{[O III]} \lesssim 4.5 \times 10^{-16}$ erg s$^{-1}$ cm$^{-2}$ \AA$^{-1}$ in the 2005 SDSS spectrum to $F_\mathrm{[O III]} = 9.8 \pm 0.5 \times 10^{-16}$ erg s$^{-1}$ cm$^{-2}$ \AA$^{-1}$ in the 2023 spectrum. These lines were detected 6 years after the initial IR flare, which, if they arise from the ionization from the TDE, implies that the gas is on the order of a few parsecs from the SMBH. Similarly fast variability in narrow lines is not that common, but it has been seen in at least one AGN with high-cadence spectral monitoring \citep{Peterson2013} as well as in extreme coronal line emitters, which are thought to be associated with TDEs in gas-rich environments \citep{Yang2013,Clark2023}. Another source that potentially shows a similar evolution is WTP\,17aamoxe, with an archival spectrum from the 6dF galaxy redshift survey \citep{Jones2004,Jones2009}. The 6dF spectrum, taken on the UK Schmidt Telescope (UKST) in 2004, does not show strong emission lines from [O III], [O I], or [S II]. However, our 2023 spectrum shows strong [O III], [O I], and [S II] features with Seyfert-like ionization levels, suggesting a recent increase in ionization levels, potentially associated with the TDE. We caution, however, that direct comparison of the narrow lines in these different spectra is complicated by the use of different telescopes, spectrographs, and seeing conditions. In particular, the 6dF survey was performed with 6.7\arcsec\, fibers, and thus any nuclear emission lines in the 2004 spectrum of WTP\,17aamoxe could be diluted by a significant contribution from the host galaxy light. Further monitoring of these sources is necessary to search for continued changes to the optical spectra.

\begin{figure*}[t!]
    \centering
    \includegraphics[width=\textwidth]{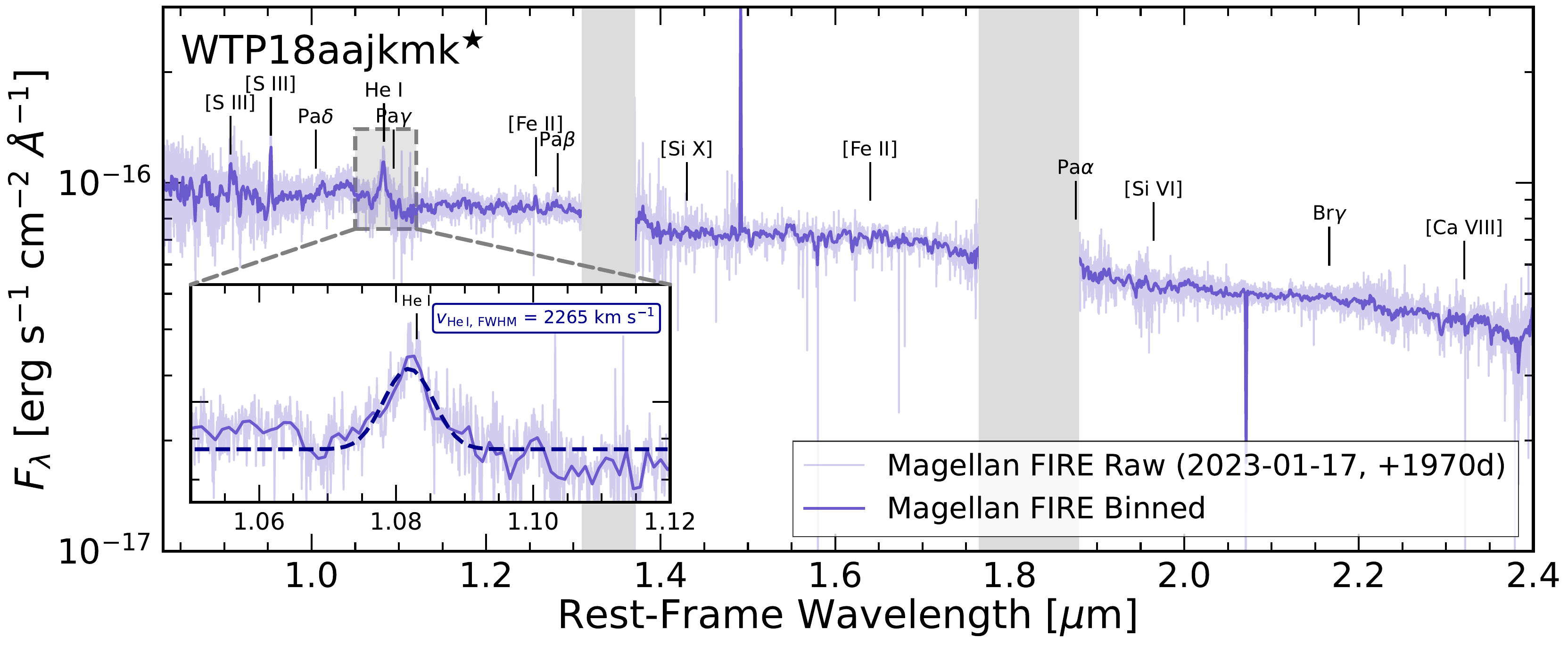}
    \includegraphics[width=\textwidth]{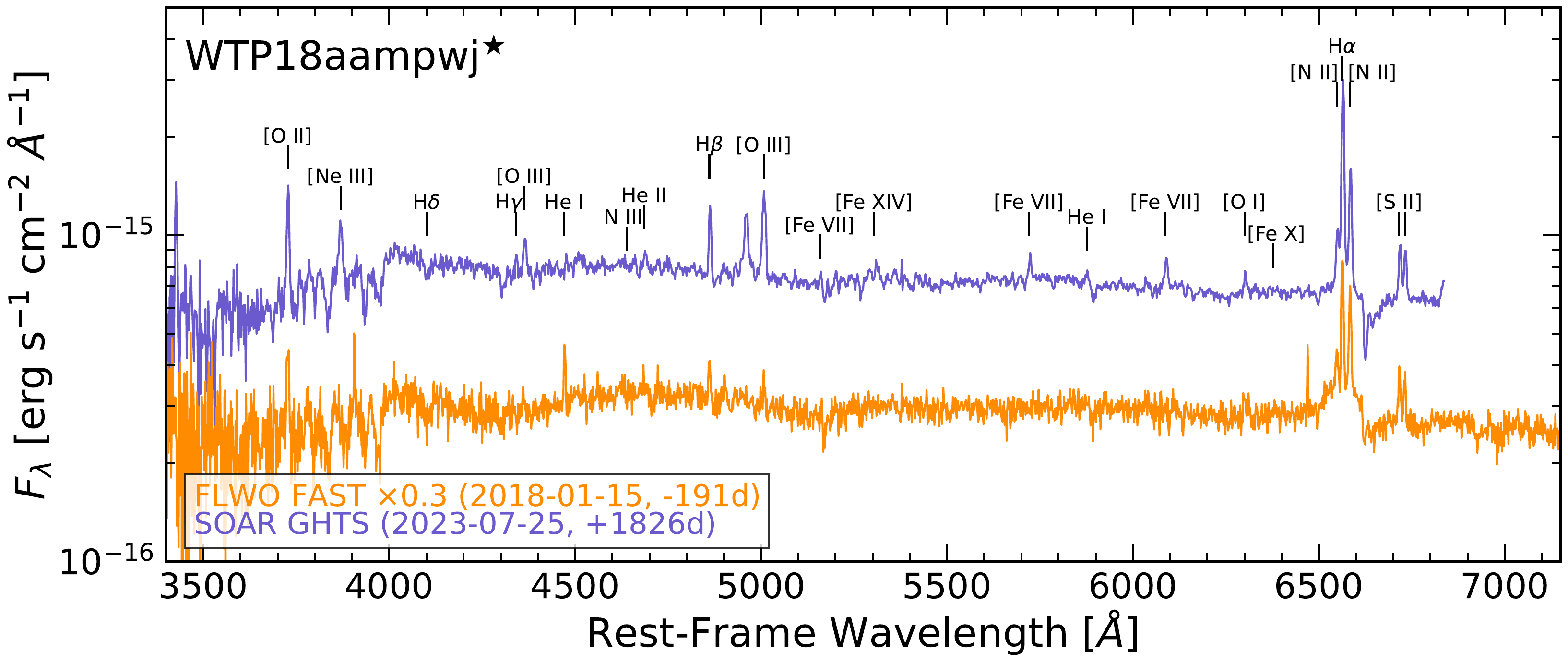}
    \caption{Evidence for transient broad emission lines in two WISE TDE candidates. \textit{Top:} near-IR spectrum of WTP\,18aajkmk taken with the FIRE instrument on the Magellan Baade Telescope in early 2023. The light purple shows the raw spectrum, which has been binned in 10 \AA\, bins in dark purple. Key near-IR lines commonly seen in AGN and TDEs are marked, although only [S III] and He I lines are significantly detected in the spectrum. Regions of telluric absorption are shaded in grey. The inset plot in the lower left shows a zoom in on the broad He I line at 1.083 $\mu$m, which has $v_\mathrm{FWHM} = 2265$ km s$^{-1}$, suggesting underlying accretion. \textit{Bottom:} Two optical spectra of WTP\,18aajkmk, the first taken with the FAST instrument at FLWO in 2018 shortly after this source was detected as an optical transient (shown in orange and offset to highlight spectral changes), and the second taken on SOAR with the GHTS instrument in July 2023 (shown in purple). Key optical emission lines in TDEs are marked. The latest spectrum shows clear evidence for increased [O III] and H$\beta$ emission, as well as narrow coronal emission lines associated with [Fe VII] at $\lambda = 5721,\, 6087$ \AA. The 2018 FLWO spectrum has a clear broad H$\alpha$ line with $v_\mathrm{FWHM} = 3900$ km s$^{-1}$, suggesting underlying accretion. The latest spectrum does not show any evidence for a broad H$\alpha$ line.}
    \label{fig:broad_lines}
\end{figure*}

Two sources in our sample show broad emission lines, which are commonly seen in optical and X-ray selected TDEs \citep[e.g.][]{vanVelzen2021c,Hammerstein2023}. Figure \ref{fig:broad_lines} shows the optical and near-IR spectra for these two sources. In the near-IR, WTP\,18aajkmk shows the appearance of a broad ($v_\mathrm{FWHM} \approx 2000$ km s$^{-1}$) He I line at 1.083 $\mu$m in 2023, which could arise from the scattering of this emission by polar dust into our line of sight \citep[e.g.][]{Kool2020}. Although we do not have archival spectroscopic observations in the near-IR of this source, the lack of other broad emission lines in the near-IR is unusual for an AGN \citep{Landt2008}. Interestingly, this spectrum was taken nearly 5 years after the initial flare, and the optical spectrum of WTP\,18aajkmk taken around the same time shows no evidence for any broad emission lines. We suggest that this could be due to decreasing flux of the broad H$\alpha$ line with time, combined with a significant amount of host galaxy H$\alpha$ light (e.g. from star formation) that can easily drown out the broadened emission. Additionally, WTP\,18aampwj, one of the few sources in our sample which has an optical counterpart in ZTF, shows a broad ($v_\mathrm{FWHM} \approx 3900$ km s$^{-1}$) H$\alpha$ line shortly after its initial detection in the optical. However, a recent spectrum taken in July 2023, reveals that the broad H$\alpha$ line has faded in the last 5 years, consistent with most optical TDEs which show fading broad lines over the course of a few years. This is also consistent with the idea that the broad Balmer lines become increasingly more difficult to detect compared to near-IR lines at late times.

The recent spectrum of WTP\,18aampwj also shows the formation of narrow coronal emission lines associated with highly ionized iron, which have also been seen in late-time spectra of two optical TDEs \citep{Onori2022,Short2023}. The most prominent lines are [Fe VII] $\lambda$5721 and [Fe VII] $\lambda$6087, which have an ionization potential (IP) of nearly 100 eV, suggesting that there must be a FUV/soft X-ray ionizing source. However, there is no evidence for higher ionization species like [Fe X] or [Fe XIV], which require IP = 235, 360 eV, respectively. Given the gap in spectral coverage from 2018-2023, we are not able to determine when the coronal lines turned on in this source, but continued monitoring will allow us to determine whether they are increasing or decreasing in flux. In addition to the coronal lines, the new spectrum of WTP\,18aampwj also shows increased strength in the various narrow forbidden lines, including [O III] and [O I], similar to WTP\,17aamoxe and WTP\,17aamzew. Thus, we suspect that the strengthening of these lines and the appearance of coronal lines are the result of reprocessing of the initial TDE flare in more distant material, akin to the narrow line region in AGN.

With the exception of WTP\,14abnpgk, which is a recent optical transient amid a secondary rise in the WISE light curves (see Section \ref{sec:appendeix_indiv} in the Appendix), the remainder of our sources show no clear evidence for transient, accretion-induced emission line changes. This could be due to the fact that many are heavily dust obscured and the majority of our follow-up efforts have been with optical spectroscopy. However, we observed six additional sources in the near-IR and found no evidence for any broad lines associated with the Paschen series or He I. Importantly, the timescales we are probing are much longer than the typical timescales studied in optical and X-ray selected TDEs, which are often tracked for up to 1-2 years. Typically, the broad lines in optical and X-ray selected TDEs fade on these 1-2 year timescales. Hence, our follow-up campaign, which ranges from 5-9 years after the IR flare, may lack spectral changes simply due to the different timescales probed. However, the appearance of accretion-induced narrow lines at $\gtrsim$ 5 years since the initial flare motivates further spectroscopic follow-up of all TDEs to longer timescales.

\subsection{Late-Time X-ray Emission}

Strong X-ray emission from the nucleus of a galaxy is an indication of ongoing accretion, either from a TDE or an underlying AGN. AGN tend to have X-ray spectra with a $\Gamma \approx 1.8-2$ power-law arising primarily from inverse Compton scattering of accretion disk photons by a hot, optically thin plasma called the corona \citep[e.g.][]{Ricci2017}. The observed spectrum is often modified by various effects, like dust and gas attenuation, disk outflows, and reflected emission, but these components can often be disentangled from one another with spectral models. TDEs, on the other hand, exhibit far less standardized X-ray emission, with some TDEs showing no X-ray emission, some showing only soft, thermal emission from the accretion disk, and others showing a late-time hardening of their X-ray spectrum, suggesting the formation of a corona \citep{Saxton2020,Guolo2023}. It is still poorly understood why there is such a range of X-ray properties in TDEs. Besides a potential dependence on the viewing angle \citep{Dai2018}, the SMBH accretion history can also influence the observed properties. Hence, understanding the X-ray emission in this sample of IR-selected TDEs is two-fold: it first provides convincing evidence for ongoing accretion in some of these systems, and second, allows us to probe the X-ray properties of a potentially new population of TDEs to compare to optically-selected samples.

\begin{figure*}[t!]
    \centering
    \includegraphics[width=\textwidth]{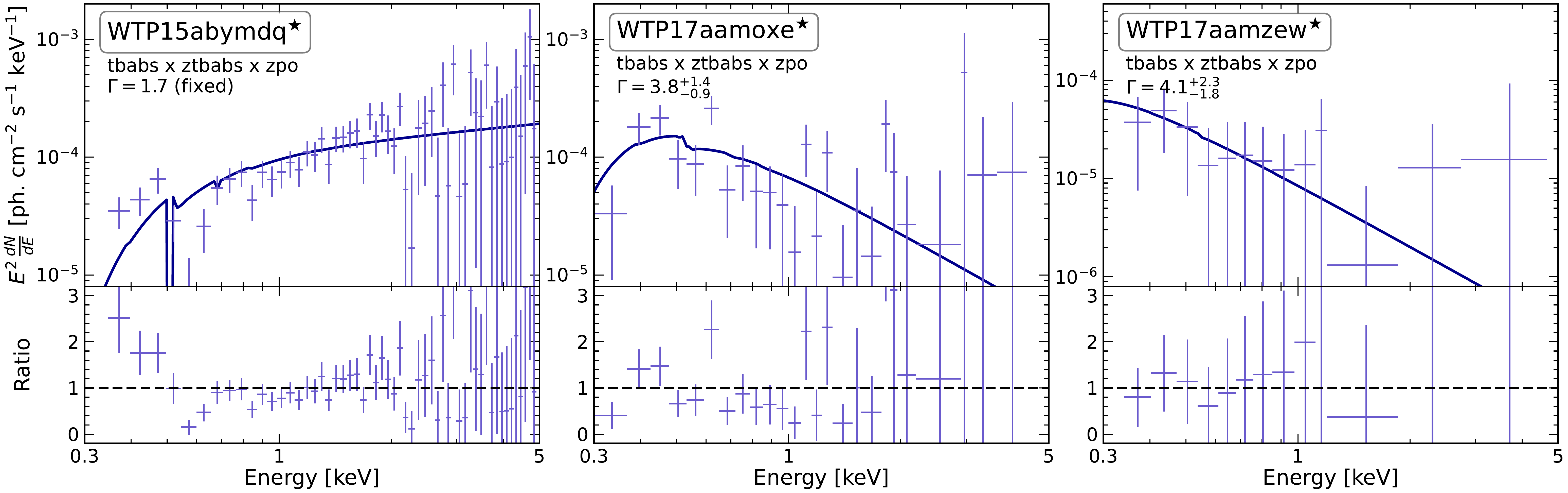}
    \caption{Spectra fit with an absorbed power law (\texttt{tbabs $\times$ ztbabs $\times$ zpo} in XSPEC notation) for the three sources that were detected in the German half of the eROSITA sky. Each panel shows a different source, where, for visual purposes, the spectrum is the combined spectrum across all eRASS detections. The best-fitting absorbed power-law model is overplotted as a solid line, where the model parameters are taken as the average of the best-fit parameters across all epochs. WTP\,15abymdq is clearly a much harder X-ray source than both WTP\,17aamoxe and WTP\,17aamzew, which are quite soft, as typically seen in many TDE X-ray spectra.}
    \label{fig:eRO_spec}
\end{figure*}

Three of the eight sources (WTP\,15abymdq, WTP\,17aamoxe, WTP\,17aamzew) in the German half of the eROSITA sky show significant X-ray detections, all in more than one eRASS epoch. All three detected sources show significant variability between the eRASS epochs, suggesting that the X-ray emission is indeed associated with accretion. As in optical and X-ray selected TDEs, their X-ray spectra show a wide range of spectral shapes and qualities. To model the eROSITA spectra, we use XSPEC \citep[v12.12.1;][]{Arnaud1996} and group the spectra with the \cite{Kaastra2016} optimal binning technique with a minimum of 1 count per bin. We fit in the 0.3-5 keV range and employ Cash statistics \citep{Cash1979} when fitting, given the low count nature of these spectra. In Figure \ref{fig:eRO_spec}, we show the best fitting eROSITA spectrum with all of the detections combined into a single spectrum and the average model parameters from all epochs. 

\subsubsection{WTP\,15abymdq}

WTP\,15abymdq was detected with eROSITA in all five eRASS epochs with a 0.2-2.3 keV luminosity ranging between $2 \times 10^{41} - 10^{42}$~erg~s$^{-1}$. The X-ray spectrum is quite hard, with a best-fit photon index ranging from $\Gamma = 0.7-2.0$ across the five eRASS epochs when no obscuration is included. Such low photon indices are often indicative of significant obscuration, and indeed, we can obtain comparable fit statistics when fitting with the photon index fixed at $\Gamma = 1.7$ and including neutral obscuration with a variable column density across the five eRASS epochs. The required column densities in each epoch span from $< 1.6 \times 10^{20}$ cm$^{-2}$ on the low end, up to $1.9 \times 10^{21}$ cm$^{-2}$ on the high end. The data can also be fit with a constant neutral and ionized absorber; in this scenario, the best fitting model has a photon index of $\Gamma = 2.0_{-0.6}^{+0.5}$ and a neutral column density of $N_\mathrm{H} < 3.0 \times 10^{20}$ cm$^{-2}$. Given the low flux of this source and the soft nature of the eROSITA response, we cannot distinguish between these two models. However, both suggest that the underlying X-ray spectrum can be a standard Comptonized emission and low levels of obscuration. Such hard X-ray emission is rather uncommon in TDEs, but has been observed in late-time X-ray follow-up of some TDEs, often when the accretion rate has dropped considerably \citep[e.g.][]{Chakraborty2021,Wevers2021,Yao2022}. Further deep monitoring in the X-ray band is needed to probe whether this source continues to be X-ray bright and AGN-like or whether the X-rays fade, as expected for a TDE when it runs out of fuel. 

\subsubsection{WTP\,17aamoxe}

WTP\,17aamoxe was detected with eROSITA in the first of four eRASS scans, with a 0.2-2.3 keV luminosity that ranged from $4 \times 10^{41} - 3 \times 10^{42}$~erg~s$^{-1}$. On the contrary to WTP\,15abymdq, WTP\,17aamoxe shows a relatively soft X-ray spectrum. When we fit all three eRASS epochs in which WTP\,17aamoxe was detected simultaneously with an absorbed power law model, the best-fit photon index is $\Gamma = 3.8_{-0.9}^{+1.4}$, with negligible neutral obscuration required. A similarly good fit can be found by assuming a blackbody model, which gives a best-fit temperature of $kT_\mathrm{bb} = 0.14 \pm 0.02$ keV. This is still relatively hot for standard accretion disks around supermassive black holes, but is comparable to many of the X-ray spectra in the eROSITA TDE sample \citep{Sazonov2021}. Allowing the temperature to vary across the observations does not provide significant improvement to the fit, but interestingly, the brightest epoch actually is slightly cooler than the two dimmer observations. Such a soft X-ray spectrum is unlike most AGN X-ray spectra \citep{Ricci2017}, but is seen rather often in optical and X-ray selected TDEs \citep{Saxton2020}.

\subsubsection{WTP\,17aamzew} \label{subsubsec:17aamzew}

WTP\,17aamzew was detected with eROSITA in the first, second, and fourth eRASS scans, with a 0.2-2.3 keV luminosity that stayed roughly constant around $2 \times 10^{41}$~erg~s$^{-1}$. However, WTP\,17aamzew was also detected by the XMM-Newton Slew Survey twice in 2018, where the X-ray luminosity was much larger ($L_{0.2-2\,\mathrm{keV}} \approx 6 \times 10^{42}$~erg~s$^{-1}$). As with the other two sources, the variable nature of the X-ray emission points to an accretion-driven origin of the X-ray emission. The eROSITA spectrum of WTP\,17aamzew is extremely soft, even more so than that of WTP\,17aamoxe. With a simultaneous fit to the three eRASS epochs in which WTP\,17aamzew was detected, we find that an absorbed power law fits well with $\Gamma = 4.1_{-1.8}^{+2.3}$ with negligible absorption required. As with WTP\,17aamoxe, the data is equally well fit by a blackbody, which gives a best fit temperature of $kT_\mathrm{bb} = 0.11_{-0.04}^{+0.06}$ keV. This is again comparable to the temperatures of other eROSITA X-ray spectra of TDEs \citep{Sazonov2021}, and relatively consistent with what is seen in TDE X-ray spectra \citep{Saxton2020}.

\section{Host Galaxy Properties} \label{sec:hosts}

\subsection{Galaxy Morphology} \label{subsec:gal_morph}

\begin{figure*}[t!]
    \centering
    \includegraphics[width=\textwidth]{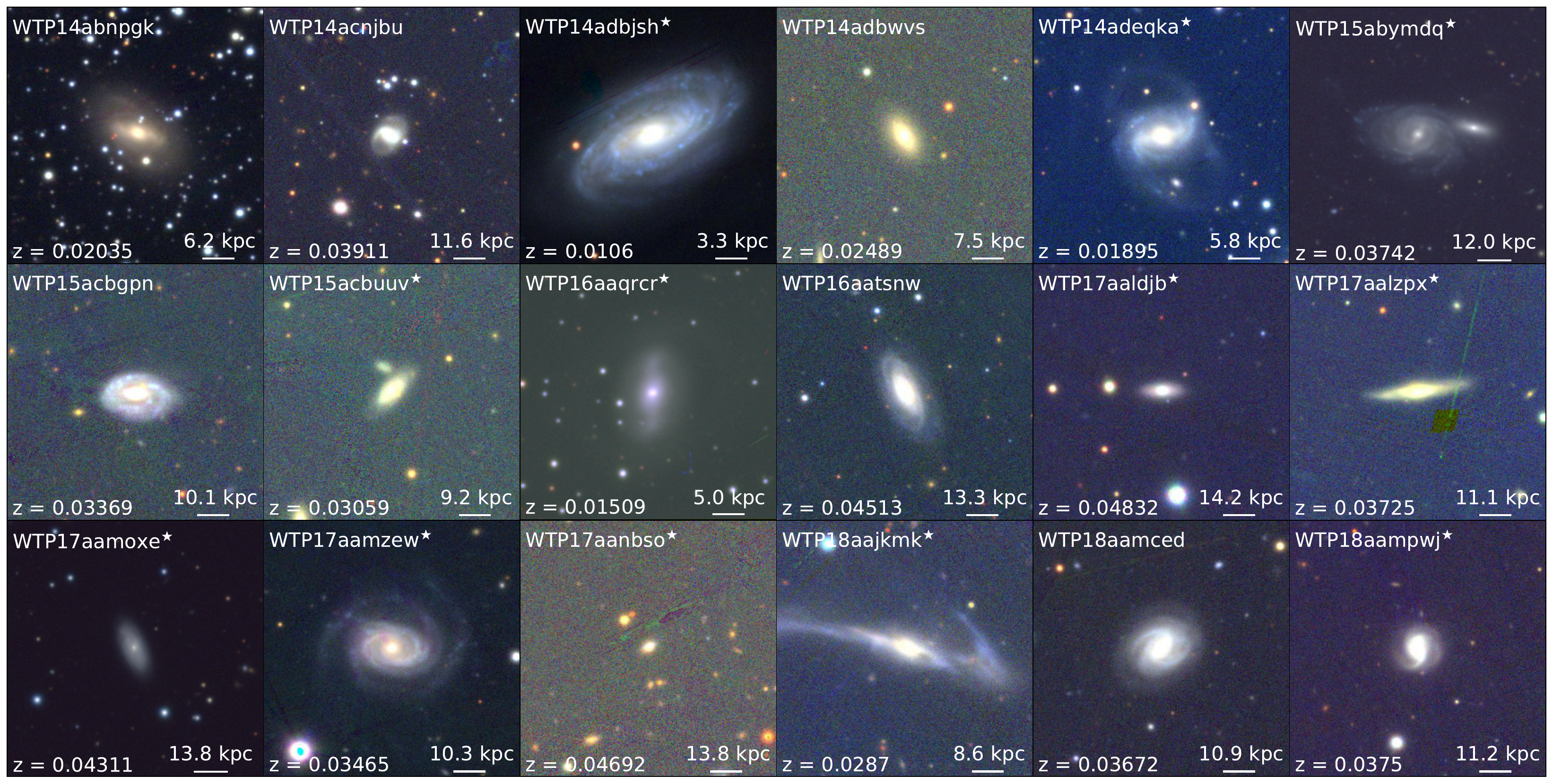}
    \caption{$gri$ cutout images (2\arcmin$\times$2\arcmin) of the host galaxy of each WISE TDE candidate. Gold sample sources are marked with~$^\bigstar$. Most images come from the PanSTARRS1 catalog, with the exception of the three most southern sources (WTP\,15abymdq, WTP\,16aaqrcr, WTP\,17aamoxe), which are taken from different surveys with DECam (DES, \citealt{Abbott2018}; MagLiteS, \citealt{Drlica-Wagner2016}). In each cutout, we list the WTP name, host redshift, and show a 15\arcsec\, scale at the bottom right.}
    \label{fig:cutouts}
\end{figure*}

In Figure \ref{fig:cutouts}, we show the host galaxy color $gri$ images of the 18 TDE candidates. There is a wide range of host galaxy morphology across the sample, including numerous galaxies with clear spiral structure and a few disturbed or merging galaxies, both of which are not usually seen in optical TDE searches. For comparison with optical TDE searches, we refer the reader to Figure 4 of \cite{vanVelzen2021c} and Figure 2 of \cite{Hammerstein2023}, which show the host galaxy color cutouts for the TDEs identified in first ZTF survey. We caution, though, that the ZTF sample and our WISE-selected sample probe very different redshift ranges, with the majority of the ZTF TDEs being found at a redshift of $z > 0.05$, whereas all of our sources are at $z < 0.05$. 

\subsection{SED Modeling} \label{subsec:seds}

For a quantitative assessment of the host galaxy properties of these TDE candidates, we model the spectral energy distribution (SED) of each host galaxy using pre-flare broad-band photometry. All photometry was obtained from existing catalogs and is corrected for Galactic extinction. We utilized archival GALEX photometry \citep{Martin2005} for both the NUV and FUV bands, including only detections. For the optical, we obtained SDSS model magnitudes \citep{Stoughton2002} where available, PanSTARRS Kron magnitudes \citep{Chambers2016} for those galaxies not in the SDSS footprint, and SkyMapper petro magnitudes \citep{Wolf2018} for the three southern-most sources, which are in neither the SDSS or PanSTARRS footprints. Finally, we include pre-flare, profile-fit photometry from the 2MASS extended source catalog \citep{Skrutskie2006} and the AllWISE catalog \citep{Wright2010} in the near- and mid-IR, respectively.

To allow for direct comparison with the literature on optical and X-ray selected TDEs, we model the SEDs using the same methodology as past work \citep[see e.g.][]{Sazonov2021,vanVelzen2021c,Hammerstein2023,Yao2023}. In short, we use the \texttt{prospector} software \citep{Johnson2021} to fit the flexible stellar population synthesis \citep[FSPS;][]{Conroy2009,Conroy2010} models assuming a delayed, exponentially declining star formation history with a Chabrier initial mass function \citep{Chabrier2003}. This model has a total of five free parameters: the galaxy stellar mass, characteristic age of the stellar population, e-folding timescale of the star formation history, metallicity, and dust optical depth, assuming the extinction law from \cite{Calzetti2000}. To estimate uncertainties on the parameters, we ran MCMC samplers using the \texttt{emcee} package \citep{Foreman-Mackey2013}. The chains contained 100 walkers and were initially populated based on a maximum likelihood fit. We utilized a conservative burn length of 500 chain steps per walker and then ran the chain for 1000 steps per walker, which was enough for all candidates to have well-sampled and stable chains. We show the resulting SED fits with the photometry in Section \ref{sec:appendix_sed} of the Appendix.

\begin{figure*}[t!]
    \centering
    \includegraphics[width=\textwidth]{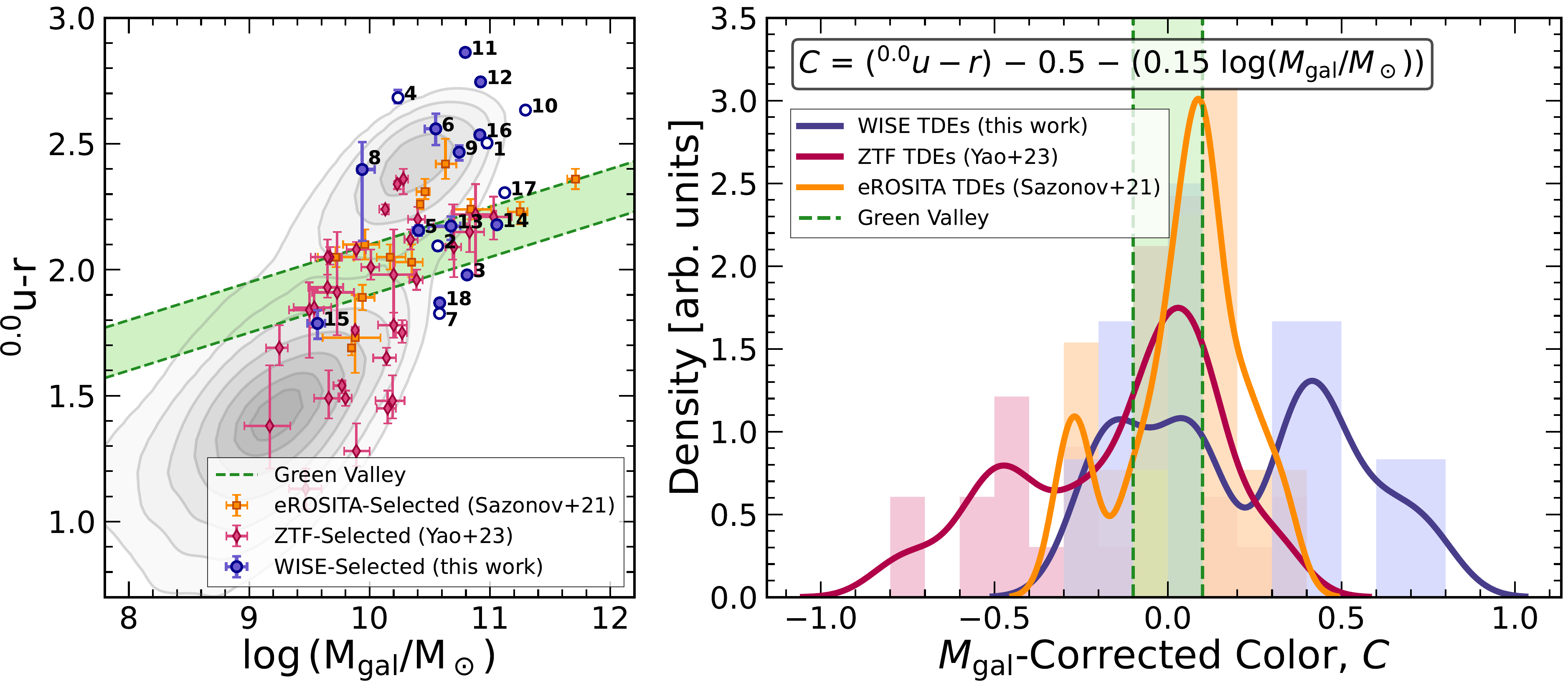}
    \caption{A comparison of the host galaxy color and mass, compared to optical and X-ray selected TDEs, indicating that the IR-selected TDEs do not show the same green valley overdensity. \textit{Left:} Rest-frame $u-r$ color of the host galaxy of TDEs versus host galaxy stellar mass. The purple circles show the WISE TDEs presented in this work, with sources in our gold sample shown as filled markers and sources in our silver sample shown with empty markers. The corresponding source ID number for each source (given in Table \ref{tab:candidates}) is shown to the top right of each point. The ZTF TDEs \citep{Yao2023} and the eROSITA TDEs \citep{Sazonov2021} are shown as red diamonds and orange squares, respectively. The green valley, as defined in \cite{Yao2023}, is shown with a shaded green region. The underlying contours are a volume-limited selection of SDSS galaxies from \cite{Mendel2014}, indicating roughly the underlying distribution of galaxies in the local universe in this color-mass space. The WISE TDEs tend to sit at higher mass than most other TDEs, which is likely the result of the relatively high limiting flux we used when selecting the sample. \textit{Right:} Histogram of mass-corrected colors, as defined in \cite{Yao2023}. The filled histograms show the raw data for the WISE (gold sample only), ZTF, and eROSITA samples, while the solid lines show the Gaussian-smoothed histograms for each sample. The location of the green valley is shaded in green with dashed lines denoting the boundaries. This highlights that the IR TDEs from this work do not show the same green valley overdensity as seen in optical and X-ray selected TDEs, favoring instead a more flat distribution in color space.}
    \label{fig:color_mass}
\end{figure*}

From the MCMC fits, we estimate the rest-frame $u-r$ color (denoted $^{0.0}u-r$) and the stellar mass of the host galaxy\footnote{For consistency with past work, we report the \textit{surviving} stellar mass (i.e. the total stellar mass times the fraction of mass which survives to present day and is not lost due to stellar winds or supernovae) for the stellar mass of the host galaxy.}. The resulting values are shown in the left panel Figure \ref{fig:color_mass}, with a sample of SDSS galaxies in the same local volume ($z < 0.05$) from \cite{Mendel2014} shown in grey contours. Past works have found that the optical and X-ray selected TDEs have a strong overdensity in the green valley \citep[e.g.][]{Sazonov2021,vanVelzen2021c,Hammerstein2023,Yao2023}, which can be seen in the optical and X-ray selected TDE populations shown in orange squares and red diamonds, respectively, in Figure \ref{fig:color_mass}. From a simple visual examination, it is evident that the IR TDEs in our sample occupy a different part of the color-mass parameters space. For a quantitative assessment, we compute the mass-corrected color, $C = (^{0.0}u-r) - 0.15 - 0.15 \log(M_\mathrm{gal}/M_\odot)$ \citep[which is the distance from the green valley, as defined in][]{Yao2023}, and compare our gold sample to the optical and X-ray selected samples using the two-sample Anderson-Darling test. We find that the optical population from ZTF \citep{Yao2023} is different from the IR-selected population with a p-value of 0.002, but the p-value between the X-ray population from eROSITA \citep{Sazonov2021} and the IR population is just 0.107. However, the combined sample of both optical and X-ray populations is different from the gold IR-selected population with a p-value of 0.002, suggesting that the IR-selected TDEs in our sample represent a different population of TDE host galaxies. In the right panel of Figure \ref{fig:color_mass}, we show the distributions of this mass-corrected color from the WISE, ZTF, and eROSITA samples. The solid lines show a Gaussian-smoothed histogram for each sample, which highlights that our IR-selected sample from WISE does not show a strong green valley overdensity and instead has a flatter distribution in color space.

Our IR-selected TDEs appear to be biased towards redder and more massive systems than the optical and X-ray selected TDEs. Naively, these findings appear inconsistent with star-forming galaxies, which are often thought of as less massive and bluer than their passive, elliptical counterparts. However, the bias towards massive systems is likely the result of only selecting the brightest IR flares due to our choice of selection criteria and the limited sensitivity of WISE. This high luminosity threshold leads to preferentially selecting more massive black holes (assuming that TDEs accrete at some roughly constant fraction of the Eddington limit) and thereby more massive host galaxies. We note that the MIRONG sample, which also uses the WISE photometry, shows a similar mass bias compared to optical TDEs \citep{Jiang2021b}, although they probe nuclear transients out to a higher redshift, which can influence the sample bias as well. Importantly, the host galaxies of the IR-selected TDEs are not too massive to host SMBHs that can disrupt Sun-like stars ($M_\mathrm{BH} \lesssim 10^8 \, M_\odot$; see Section \ref{subsec:disc_energy} for approximate SMBH mass estimates from scaling relations using host galaxy mass). 

Moreover, the redder colors of the host galaxies of the WISE TDEs could be due either to heavy dust reddening or older stellar populations. Disentangling these two factors requires more sophisticated modeling, as the current single dust attenuation screen and simple parametric star formation history predicts extremely high ages (near the maximum allowed value of 12.5 Gyr for the majority of the sources). We investigated the correlation between these two parameters in our SED modeling by fixing the age at 6 Gyr and found the same qualitative results that the IR-selected TDEs are generally redder and more massive. This did lead to a slight decrease in the masses (0.15 dex on average), but produced consistent colors on average. We also tried fixing the dust optical depth at the value that can be derived using the Balmer decrement from the optical spectra. These spectra are taken from the nucleus, however, and yielded poor fits to the photometric data of the host galaxy as a whole. This suggests that the dust content of these galaxies is highly spatially variable and that more sophisticated modeling of the dust, rather than a simple single screen, is necessary to learn more about more about the host galaxies from their SEDs. This is beyond the scope of this initial sample paper, but will be explored in follow-up work.

\subsection{Spectroscopic Classification} \label{subsec:hdelta}

In addition to a green valley overdensity, optically-selected TDEs have also been found to preferentially occur in quiescent Balmer-strong, E+A, or post-starburst galaxies \citep[e.g.][]{Arcavi2014,French2016,Law-Smith2017,Graur2018,Hammerstein2021}. These systems are characterized by low H$\alpha$ equivalent width (EW), which indicates very little active star formation, coupled with a strong H$\delta_\mathrm{A}$ absorption index, which traces the amount of A type stars and thus indicates significant star formation in the last 1 Gyr. \cite{French2016} found that the majority of optical TDEs from SDSS, ASAS-SN, Palomar Transient Facility, and PanSTARRS fell into this class of post-starburst galaxies. Recent works with larger samples of both optical and X-ray selected TDEs, albeit with somewhat more relaxed criteria for what qualifies as a TDE, have found that this overrepresentation exists but is not as strong as in the initial sample \citep{Law-Smith2017,Graur2018,Hammerstein2021}. As with the green valley overdensity of TDE hosts, the prevalence of post-starburst systems has been potentially linked to a temporary enhancement in the TDE rate in post-merger systems, which tend to be more centrally concentrated.

For those sources in our sample with coverage in the H$\delta$ part of the optical spectrum ($\lambda$ = 4102 \AA), we compute the H$\alpha$ EW and the H$\delta_\mathrm{A}$ absorption index to compare to optical and X-ray selected TDEs. Following past studies, we remove the stellar absorption contribution in the H$\alpha$ region before fitting for the EW (see Section \ref{subsec:agn_contam} for more details on the removal of the stellar continuum). For those sources in SDSS, we adopt the H$\delta_\mathrm{A}$ absorption indices from the MPA-JHU catalog \citep{Brinchmann2004}. We also utilize the SOAR GHTS blue-sided spectrum of WTP\,14adbjsh, the Gemini GMOS observation of WTP\,17aamoxe, and the archival FLWO FAST observation of WTP\,18aampwj, both of which have spectral coverage in the H$\delta$ portion of the spectrum. For these objects, we follow the procedure and bandpasses outlined in \cite{Worthey1997} for computing H$\delta_\mathrm{A}$. We caution, however, that for sources in the silver sample or sources with spectral measurements from more than 5 years after the IR flare (e.g. WTP17aamoxe, \#13), there may be contamination of the H$\alpha$ line from accretion-driven excitation, thereby complicating a direct use of H$\alpha$ as a proxy for star-formation.

\begin{figure*}[t!]
    \centering
    \includegraphics[width=0.95\textwidth]{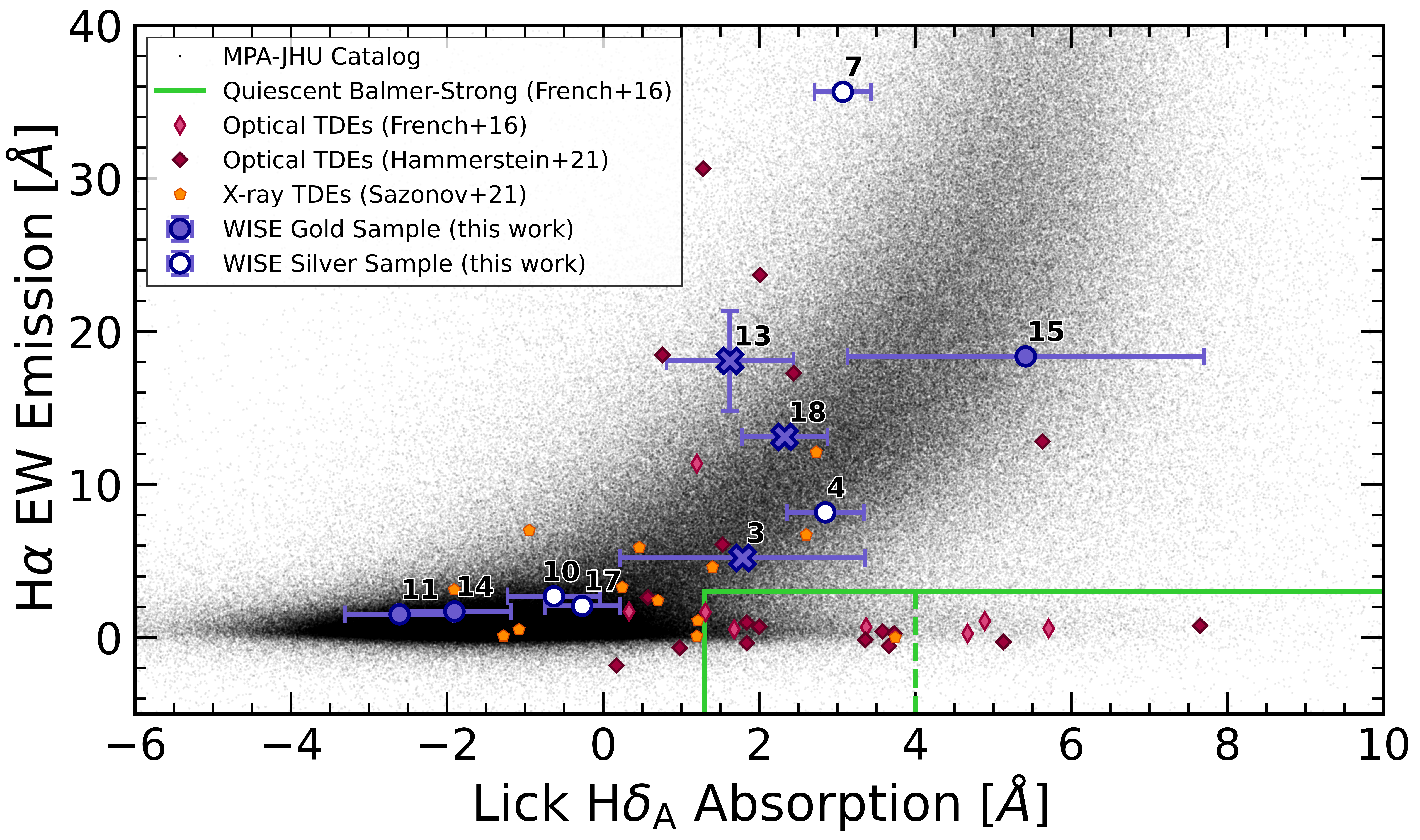}
    \caption{Equivalent width of the H$\alpha$ line versus the Lick H$\delta_\mathrm{A}$ absorption index for the 10 sources in our sample with spectral coverage around the H$\delta$ line (6/12 gold sources, 4/6 silver sources). The purple circles show data from SDSS, while $\times$'s show data from other spectrographs. Sources in our gold sample are shown with filled markers, while sources in our silver sample are shown with empty markers. All of the WISE TDEs are also labeled to the upper right with the ID number from Table \ref{tab:candidates}. Error bars represent 1$\sigma$ uncertainty. Other optical studies of the post-starburst nature of TDEs are shown with red diamonds \citep{French2016,Hammerstein2021}, and the X-ray selected TDEs from eROSITA are shown with orange pentagons \citep{Sazonov2021}. The black data points are from the SDSS MPA-JHU catalog \citep{Brinchmann2004}. Finally, the green lines show the designation of quiescent Balmer-strong (or E+A) galaxies from \cite{French2016}, whereby H$\delta_\mathrm{A} > 1.31$ \AA\, for weak classification and H$\delta_\mathrm{A} - \sigma(\mathrm{H}\delta_\mathrm{A}) > 4$ \AA\, for strong classification. None of the sources in our sample fall into this regime and the majority of our sources have significant H$\alpha$ emission, indicating that we have found TDEs primarily in star-forming galaxies.}
    \label{fig:lickHd}
\end{figure*}

None of our sources sit in the post-starburst regime, defined by \cite{French2016} as having H$\alpha$ EW $<$ 3 \AA\, and either H$\delta_\mathrm{A} > 1.3$ \AA\, or H$\delta_\mathrm{A} - \sigma(\mathrm{H}\delta_\mathrm{A}) > 4.0$ \AA, for the weak and strong post-starburst classification respectively. We show the 10 sources in our sample with H$\delta_\mathrm{A}$ measurements in Figure \ref{fig:lickHd}, along with the optical and X-ray selected TDEs for comparison. While the remaining 8 sources in our sample do not have sufficient coverage to allow us to compute the H$\delta_\mathrm{A}$ absorption index, many of these sources show strong H$\alpha$ emission EW, indicating that there is active star formation going on in many of these systems. To test whether the 6 gold sources were drawn from the underlying population, we utilized a 2D implementation of the Kolmogorov-Smirnov (KS) test, with re-sampling of the underlying MPA-JHU sample to compute the p-value. We find a p-value of 0.52, indicating that the WISE TDE sample is consistent with arising from the full MPA-JHU catalog. Hence, this IR-selected TDE sample does not have the same overrepresentation of post-starburst galaxies compared to optically-selected populations. 

\section{A Mid-IR View of TDE Rates} \label{sec:rates}

This new population of IR-selected TDEs seems to be hidden from typical detection methods, like optical and X-ray surveys, and has been unaccounted thus far in TDE demographics. Therefore, this sample has important implications for the overall rate of TDEs in the local universe, which we estimate in this section. For this, we utilize only the gold sample, which was defined in Section \ref{subsec:clean_sample} and is given in Table \ref{tab:sampledesignation}.

The observed rate of TDEs can be easily computed, using
\begin{equation}
    \mathcal{R}_\mathrm{obs} = \frac{N_\mathrm{IR \, TDEs}}{N_\mathrm{CLU}} \times \frac{1}{t_\mathrm{samp}}
\end{equation}
where $N_\mathrm{IR \, TDEs}$ is the number of TDEs we detected in our sample (12 for the gold sample), $N_\mathrm{CLU}$ is the total number of galaxies we cross-matched the WISE sample against ($2.72 \times 10^5$ for the CLU sample), and $t_\mathrm{samp}$ is the time period we are sensitive to with our search (5 years for the case of this survey). This yields a rate for our gold sample of $8.8 \pm 2.6 \times 10^{-6}$ galaxy$^{-1}$ year$^{-1}$. However, this simplistic rate fails to account for numerous biases, including the fact that galaxy samples are incomplete at even the relatively low redshifts probed in this sample \citep{Fremling2020} and that the cadence and depth of the WISE survey means we may have missed some transients. 

We account for missing events due to the cadence and depth of the WISE survey by computing the detection efficiency (DE) using simulated light curves. For each source, we simulated 10$^3$ light curves based on the best-fitting dust model detailed in Section \ref{subsec:dust_dist}. We then computed the probability that the source was detected in each of the simulations by applying the following detection criteria: the source must be detected at a peak IR luminosity of $L_\mathrm{W2} > 10^{42}$~erg~s$^{-1}$, be detected in difference imaging for at least 5 WISE epochs ($\sim$2.5 years), and not saturate the WISE detectors. The Malmquist bias is accounted for in these simulations by allowing the redshift of each simulated light curve to vary, assuming a uniform number density, such that the number of sources scales as with $d_L^3$ due to the increasing volume. The majority of the sources in our sample have a detection efficiency that is very close to 1, indicating that we detected most of the WISE TDEs with $L_\mathrm{W2} > 10^{42}$~erg~s$^{-1}$ in the galaxies in this sample. Only the sources with the lowest peak luminosity have significantly lower detection efficiencies, which primarily arise because they would not be detected in 5 WISE epochs. 

To account for the incompleteness of the CLU galaxy catalog, we utilized the redshift completeness fraction (RCF) from \cite{Fremling2020}, which estimates the probability that a given nearby galaxy will have a spectroscopic redshift. The RCF is a function of both redshift and the W1 absolute magnitude (i.e. a proxy for stellar mass). Hence, as with the detection efficiency, we compute the RCF for each of our TDEs with their respective redshift and W1 magnitude.

The detection efficiency and galaxy completeness factors are then used as weights for each TDE in our sample. The true rate is then given by
\begin{equation}
    \mathcal{R} = \frac{1}{N_\mathrm{CLU}} \times \frac{1}{t_\mathrm{samp}} \times\sum_i \left(\frac{1}{\mathrm{RCF}_i} \times \frac{1}{\mathrm{DE}_i}\right),
\end{equation}
where $\mathrm{RCF}_i = \mathrm{RCF}\left(z_i, M_{W1,\,i}\right)$ and $\mathrm{DE}_i = \mathrm{DE}\left(L_{\mathrm{IR}, i}\right)$. This yields a total rate of mid-IR TDEs of $2.0 \pm 0.3 \times 10^{-5}$ galaxy$^{-1}$ year$^{-1}$, where the uncertainty here accounts for the uncertainty on both the detection efficiency and galaxy completeness, as well as the Poisson uncertainty. The detection efficiency and galaxy completeness contribute roughly equally to the total correction to the rate, and the dominant source of uncertainty is the Poisson uncertainty, given the relatively low number of sources in our sample. The corresponding volumetric rate of IR TDEs of $1.3 \pm 0.2 \times 10^{-7}$ Mpc$^{-3}$ year$^{-1}$, assuming we probe out to $d_L = 215$ Mpc. We caution that the rate derived for this sample should be interpreted as a lower limit on the total number of IR TDEs, as our sample is not complete and may miss sources with lower covering factors (and therefore lower IR luminosities), shorter IR flares, and the rare, highly luminous flares that require a larger volume to be probed.

\begin{figure}[t!]
    \centering
    \includegraphics[width=0.46\textwidth]{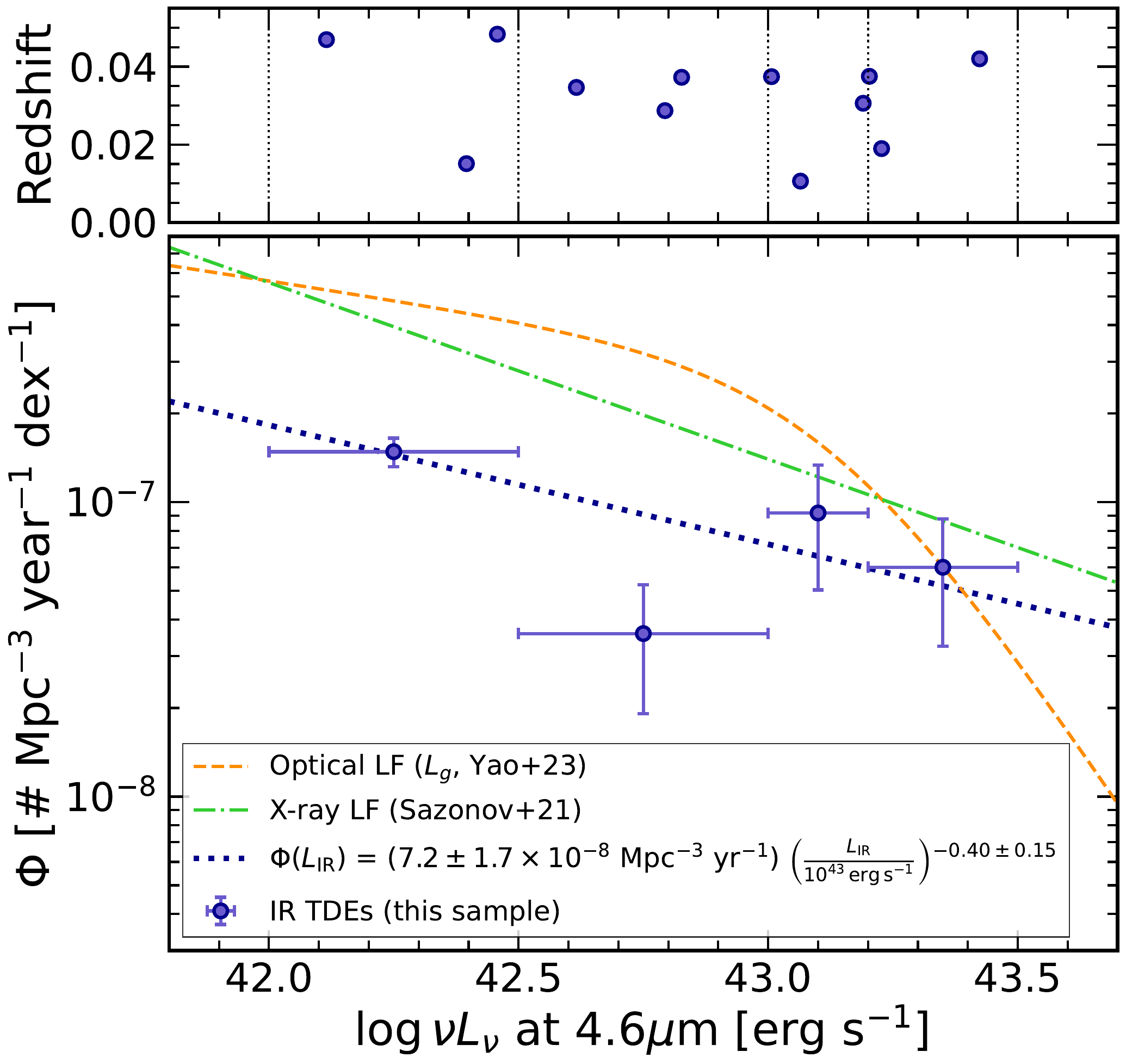}
    \caption{\textit{Top:} Redshift of our IR TDE sample as a function of peak W2 luminosity. The vertical dotted lines show the boundaries of the luminosity bins used below, chosen such that each bin contains 3 sources. \textit{Bottom:} Luminosity function (LF) of our IR TDE sample in terms of the peak W2 luminosity. This was computed using only the 12 gold tier events, given in Table \ref{tab:sampledesignation}. The purple dotted line shows the best fitting single power-law model for the IR TDE sample. For comparison, we show the optical LF for the $g$-band luminosity in orange from \cite{Yao2023} and the X-ray LF in green from \cite{Sazonov2021}. We caution, though, that a direct comparison between these samples depends on their respective luminosities, which for the IR TDE sample requires assumptions made about the covering factor.}
    \label{fig:lum_fun}
\end{figure}

In Figure \ref{fig:lum_fun}, we show the corresponding luminosity function (LF) for the IR TDE sample and compare to the optical and X-ray selected LFs presented in \cite{Yao2023} and \cite{Sazonov2021}, respectively. We find that the IR LF show tentative evidence for a decreasing rate with increasing luminosity, similar to what is found in optical and X-ray selected samples. When fit with a single power law model, given by
\begin{equation}
    \Phi(L_\mathrm{IR}) = \dot{N}_0 \left(\frac{L_\mathrm{IR}}{10^{43} \, \mathrm{erg} \, \mathrm{s}^{-1}}\right)^{-\gamma},
\end{equation}
we find $\dot{N}_0 = 7.2 \pm 1.7 \times 10^{-8}$ Mpc$^{-3}$ year$^{-1}$ and $\gamma = 0.40 \pm 0.15$. The power law index is somewhat shallower than the optical LF from ZTF and X-ray LF from eROSITA, likely due to our relatively small sample volume. Our small sample volumes means that we only probe up to a moderate peak W2 luminosity of $\log L_\mathrm{W2} \approx 43.4$. Therefore, we do not capture the relatively rare, luminous events that constrain the high end of the TDE LF, and hence, we cannot see the significant rate suppression at high luminosities due to stars being captured before disruption around massive black holes \citep[$M \gtrsim 10^8 \, M_\odot$; e.g.][]{vanVelzen2018}. Both the optical and X-ray samples probe out to significantly higher redshifts \citep[$z \sim 0.6$;][]{Sazonov2021,Yao2023}, hence hampering our ability to directly compare the LFs. Likewise, a direct comparison of the different LFs is affected by the different luminosities probed with different selection methods, especially as the observed IR luminosity is directly related to the covering factor and may underestimate the intrinsic flare luminosity. If the covering factor is anything less than 1, then the intrinsic luminosities are indeed higher than what we observe, shifting the LF to the higher luminosities. However, the lack of a sharp turn over in the observed IR LF implies that the covering factor cannot be too small ($f_c \gtrsim 10\%$), as we do not see a suppression in the high luminosity rate from massive black holes. This is consistent with the findings from the dust modeling presented in Section \ref{subsec:dust_dist}.

\section{Discussion} \label{sec:disc}

\subsection{Implications for TDE Host Galaxies} \label{subsec:disc_hosts}

One of the main drivers for conducting this survey was to look for TDEs in dusty, star-forming galaxies, which appear to be underrepresented when using optical and X-ray detection methods. Interestingly, the majority of the host galaxies in our sample show spiral morphology in their color cutouts, but have $u-r$ colors which are redder than the typical blue cloud. However, while elliptical galaxies almost exclusively occupy the red cloud in color-mass space, spiral galaxies are known to show a wide range of colors, including both the red cloud and blue sequence in color-mass space \citep[see Figures 8 and 12 of][]{Blanton2009}. In particular, the more bulge-dominated and earlier type spirals tend to show significantly redder colors and be more massive than their later type counterparts. Hence, the color-mass diagram alone is insufficient for diagnosing the nature of these host galaxies; the red colors and high masses we find do not rule out a star-forming nature of the host galaxies of these IR TDEs. The morphology and the significant H$\alpha$ emission seen in the majority of sources actually suggest that the majority of our host galaxies are indeed star-forming systems. 

We also do not see a comparable overdensity in the green valley of color-mass space compared to optical and X-ray searches. However, given the complexities of interpreting the color-mass location, especially when dealing with star-forming systems, we also approached the nature of the IR TDE hosts from a spectroscopic standpoint. Namely, the IR-selected TDEs do not appear to be concentrated in post-starburst or quiescent, Balmer-strong galaxies, which was previously found for the host galaxies of optical TDEs \citep[e.g.][]{French2016,Hammerstein2021}. Instead the IR-selected TDE sample appears to be more representative of the underlying population of SDSS galaxies, suggesting that the overabundance of optical TDEs in post-starburst is at least partially driven by selection effects. This suggests that the use of host galaxy properties to select TDEs requires caution and may miss dusty TDEs in particular.

\subsection{Implications for TDE Rates} \label{subsec:disc_rates}

One of the holy grails in the field of TDEs is to use the observed TDE LF to measure fundamental properties of black holes, like mass and spin \citep{Kesden2012}. However, there are known caveats and complications to this, including degeneracies in the LF with the age of the stellar populations \citep{Huang2024} and delayed disk formation \citep{Wong2022}. It is thus crucial to probe the entire population of TDEs to understand the underlying LF and TDE properties. Therefore, by finding a new population of TDEs that have different host galaxy properties and also probe the high end of the mass distribution, we become closer to achieving the ultimate goal of using TDE rates and LFs to probe black hole demographics.

In Section \ref{sec:rates}, we derived a rate of IR TDEs of $2.0 \pm 0.3 \times 10^{-5}$ galaxy$^{-1}$ year$^{-1}$, with a corresponding volumetric rate of $1.3 \pm 0.2 \times 10^{-7}$ Mpc$^{-3}$ year$^{-1}$. These rates are comparable, but slightly smaller than those from optical and X-ray TDE samples, with the most recent flux-limited sample of ZTF TDEs suggesting rates of $3.2_{-1.2}^{+1.8} \times 10^{-5}$ galaxy$^{-1}$ year$^{-1}$ ($3.1_{-1.0}^{+0.6} \times 10^{-7}$ Mpc$^{-3}$ year$^{-1}$; \citealt{Yao2023}) and the eROSITA sample of TDEs given rates of $1.1 \pm 0.5 \times 10^{-5}$ galaxy$^{-1}$ year$^{-1}$ ($2.1 \pm 1.0 \times 10^{-7}$ Mpc$^{-3}$ year$^{-1}$; \citealt{Sazonov2021}). Given that most optical TDEs are X-ray faint, most X-ray TDEs are optically faint, and the majority of our sample shows no transient optical or X-ray flare, the total TDE rate is likely close to the sum of these three rates. This implies a total TDE rate on the order of $6.3 \pm 1.6 \times 10^{-5}$ galaxy$^{-1}$ year$^{-1}$, which is much closer to the typical theoretical prediction of $10^{-4}$ galaxy$^{-1}$ year$^{-1}$ \citep{Wang2004,Stone2016}, although there are also sizable uncertainties in the predicted rate \citep[e.g.][]{Pfister2020}. 

It is important to note that the IR rates from our sample are likely an underestimate of the true IR TDE rate. In particular, we are sensitive only to relatively bright IR echoes ($L_\mathrm{W2} \gtrsim 10^{42}$~erg~s$^{-1}$) and require the TDEs in our sample to be long-lived ($t_\mathrm{duration} \gtrsim$ 2.5 years). Both of these points are required given the limited cadence and depth of WISE, in an effort to minimize sample contamination from SNe. Dust echoes around optical and X-ray TDEs are often shorter-lived and dimmer, likely due to a lower dust covering factor than what we see in our sources \citep[for a recent comprehensive analysis of optical TDE dust echoes see][]{Jiang2021a}. It is possible that TDEs with moderate dust covering factors are missed by both surveys, evading our selection by having low luminosity and hence not being detectable for as long. Future time-domain IR surveys with deeper limits and higher cadence are necessary to search for this population.

The rate we derive for IR-selected TDEs is smaller than that of the MIRONG sample by a factor of $\sim 2.5$ \citep[$\mathcal{R}_\mathrm{MIRONG} \approx 5.4 \times 10^{-5}$ galaxy$^{-1}$ year$^{-1}$;][]{Jiang2021b}. We suspect that this discrepancy is due to the different selection criteria of the MIRONG sample, which only required a brightening of 0.5 mag in either W1 or W2. This allowed for a number of sources in their sample that were dimmer than our peak luminosity requirement, which could contribute to the difference in rates. Likewise, as the MIRONG sample was agnostic to underlying AGN activity, their rate may be higher due to contamination from AGN. Interestingly, when we include the silver tier events in our sample, we find a rate of $4.7 \pm 1.2 \times 10^{-5}$ galaxy$^{-1}$ year$^{-1}$, which is consistent with the MIRONG rate within uncertainties. Even this silver tier sample has some measure of filtering for AGN activity (WISE colors, no extraneous WISE variability), unlike the MIRONG sample, which could imply that the consistency with the MIRONG rate is due to difference imaging techniques revealing more events than the standard PSF photometry techniques used in the WISE catalogs.

\subsection{IR TDEs \& The Missing Energy Problem} \label{subsec:disc_energy}

Another long-standing discrepancy among TDE observations and theory concerns the total energy budget of TDEs. If roughly half of a Sun-like star becomes bound and is accreted onto the black hole when a TDE occurs, then the expected energy released should total around $10^{52-53}$ erg. Yet, usually only about 1\% ($\sim 10^{51}$ erg) of this emission is seen in either optical or X-ray TDE searches. There have been a number of proposed solutions to this problem, including photon trapping or significant outflows in super-Eddington accretion flows at early times \citep{Begelman1979,Abramowicz1988,Strubbe2009,Coughlin2014,Metzger2016}, low radiative efficiency in eccentric accretion flows \citep[e.g.][]{Svirski2017}, and off-axis relativistic jets \citep[although this scenario is disfavored based on radio observations of non-relativistic TDEs; e.g.][]{Lu2018}. 

Another suggestion for solving the missing energy problem is to have the majority of the radiation released in the extreme UV (EUV), which is inaccessible to our current suite of telescopes \citep{Lu2018}. This has been supported by recent simulations of super-Eddington accretion flows from TDEs, where when the output spectra were fit with a blackbody in either the X-ray or optical, only 1-10\% of the total output energy was captured \citep{Thomsen2022}. Dust, however, is an efficient reprocessor of EUV radiation, and hence, looking at the output energy in the IR can help us probe the underlying bolometric energy output of TDEs to test this theory. Past work on individual IR flares in ultra-luminous IR galaxies (ULIRGs) have found some evidence for this missing energy, with more than $10^{52}$ erg radiated during these outburst in the IR alone \citep{Dou2017,Tadhunter2017,Mattila2018}, although both of these sources show evidence for pre-flare AGN activity. 

\begin{figure}[t!]
    \centering
    \includegraphics[width=0.48\textwidth]{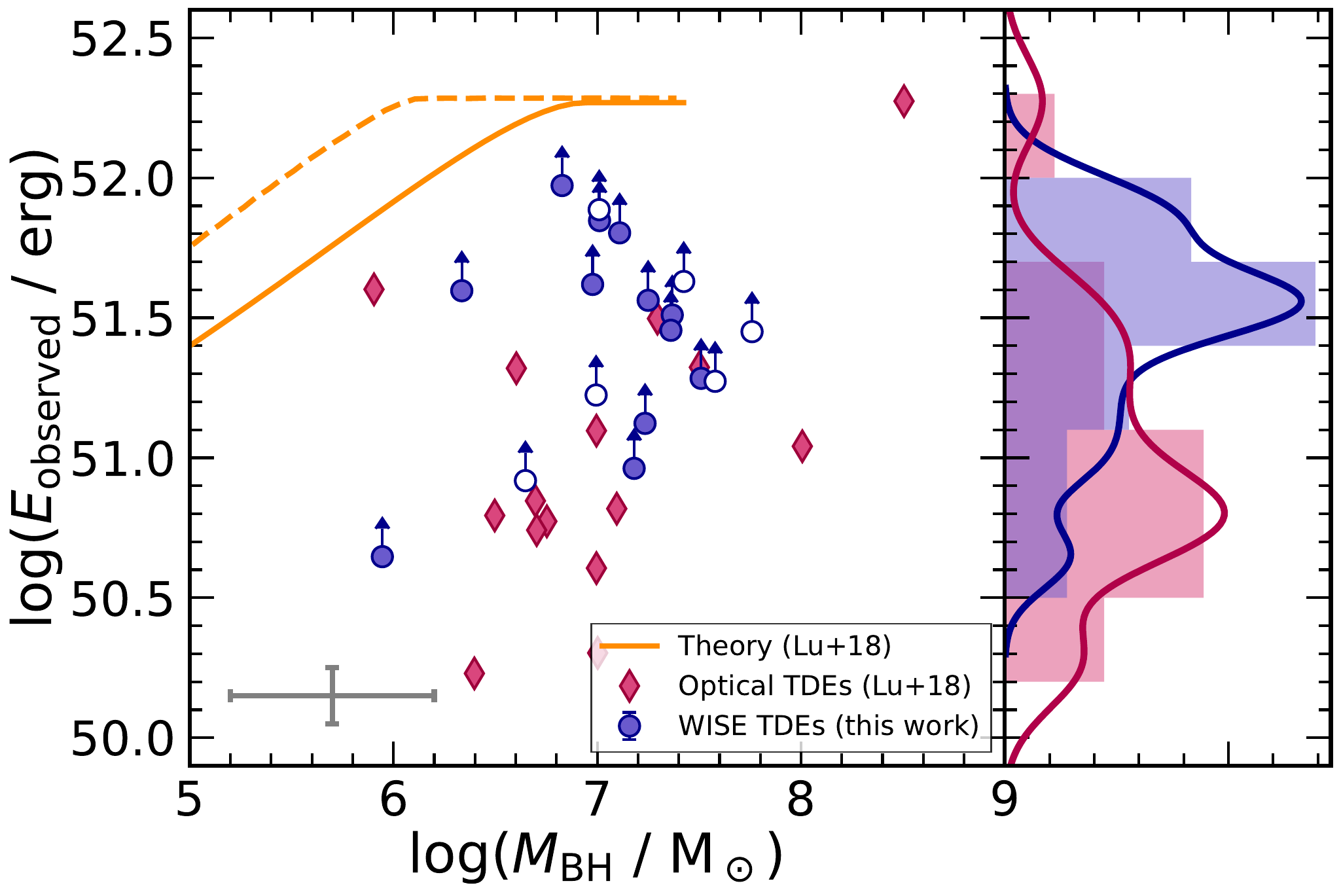}
    \caption{Illustration of the missing energy problem in TDEs, highlighting that the IR TDEs selected in this sample bring the observed values closer to the theoretical expectation. \textit{Left:} Observed radiated energy versus black hole mass. The purple circles show the integrated IR energy reported in Table \ref{tab:sampledesignation}, where filled circles are from the gold sample and empty circles are from the silver sample. These points do not account for the covering factor and are therefore lower limits on the total radiated energy. The black hole masses for these points are estimated using the \cite{Reines2015} scaling relation between galaxy stellar mass and black hole mass. A representative error bar is shown in grey in the bottom left. The pink diamonds are the optical TDEs given in \cite{Lu2018}, and the two orange lines denote theoretical expectations from \cite{Lu2018} for the minimum radiated energy in a TDE. The dashed and solid lines correspond to different model assumptions about the specific energy and disrupted star, representing the most (frozen, $\gamma = 4/3$) and least (flat, $\eta_\epsilon = 1$) extreme cases from \cite{Lu2018}, respectively. \textit{Right:} Histogram of the two samples, with WISE data in purple and optical data in pink. The solid histograms show the raw data and the lines show a smoothed histogram.}
    \label{fig:missing_energy}
\end{figure}

The integrated energy in the IR for our sources totals between $4 \times 10^{50} - 9 \times 10^{51}$ erg, which is shown in Figure \ref{fig:missing_energy} versus black hole mass \citep[estimated using the scaling relation with galaxy stellar mass from][]{Reines2015}. These energies are slightly larger than the output in the optical, UV, and X-ray TDEs, but are not large enough to immediately resolve the missing energy puzzle. However, they should be considered a lower bound on the total radiated energy, as a covering factor less than 1 reduces the observed luminosity. While we cannot accurately estimate the covering factor with the lack of optical flares, we constrain the covering factor to be on the order of $f_c \gtrsim 10\%$, which can allow the integrated IR energy to be somewhere between the observed value and one order of magnitude higher. For the majority of the sources in our sample, this is comparable with the expected 10$^{52-53}$ erg, although there are a few sources that would still be lower than this theoretical expectation. It is also important to note that our sample only probes an extremely small volume ($z \lesssim 0.05$) and therefore misses many of the rare, higher luminosity IR TDEs. This motivates further studies of IR TDEs to larger volumes to probe the missing energy problem in the most luminous IR events. We will be able to probe larger volumes with WISE in future studies, as our current sample is limited by our choice of galaxy survey for cross-matching, rather than the WISE sensitivity. In summary, our nearby sample pushes closer to the expected theoretical values for the total radiated energy in TDEs and allows for a comparison to samples of nearby optical and X-ray selected TDEs.

\subsection{Comparison to Optical and X-ray Selected TDEs}

Numerous optical and X-ray selected TDEs have shown dust echoes detected with the WISE satellite \citep{Dou2016,Jiang2016,vanVelzen2016,Jiang2021a,Sazonov2021,Hinkle2022}. A comprehensive analysis of 10 years of optical TDEs by \cite{Jiang2021a} revealed that the TDEs with detected dust echoes had typical covering factors of only $f_c \approx 1\%$ and typical peak IR luminosities of only $L_\mathrm{dust} \approx 10^{41-42}$~erg~s$^{-1}$. By definition, our sample includes only brighter IR echoes than this, hence why our sample contains no previously-known optical TDEs. Through dust modeling, we found that the covering factor in these sources is $f_c \gtrsim 10\%$, significantly higher than the 1\% number derived for optical TDEs. As expected, this implies that the IR TDEs originate in much dustier environments than optical TDEs. Probing the existence of an intermediate sample of TDEs with covering factors on the order of $f_c \approx 1-10\%$ would require different sample selection criteria, namely not requiring such large luminosities and lengthy flares. 

Interestingly, the dust echo in the sample of optical TDEs which resembles our sample the most is ASASSN-15lh, which has the longest-lived dust echo of any of the optical TDEs \citep{Jiang2021a}. The nature of this source has been highly debated, with both a SN origin \citep[e.g.][]{Dong2016,Godoy-Rivera2017} and a TDE origin \citep[e.g.][]{Leloudas2016,Kruhler2018,Mummery2020} being argued. The primary difference between ASASSN-15lh and our sample is the strong optical flare seen in ASASSN-15lh; while our sources show a strong IR flare with either no or very weak optical flares, ASASSN-15lh shows a strong optical flare with $L_\mathrm{opt} \approx 10^{45}$, still suggesting a relatively low covering factor like other optical TDEs. 

Our findings suggest that the IR TDEs we have found are in dust-rich galactic nuclei. Similarly, transient coronal lines have recently been seen in numerous sources, and have been suggested to be light echoes of TDEs in gas-rich environments \citep[e.g.][]{Komossa2008,Wang2012,Clark2023,Hinkle2023}. Coronal lines are high ionization lines (ionization potentials (IP) $\gtrsim$ 100 eV), which were named such due to their existence in the hot solar corona. Interestingly, \cite{Clark2023} recently studied the long-term evolution of some extreme coronal line emitters (ECLEs) and found that the mid-IR WISE light curves showed long and steady declines, similar to what we see in our WISE TDE sample. Additionally, some optical/UV TDEs (and one ambiguous nuclear transient, AT\,2019avd) have shown transient coronal lines, and all of these sources show a strong IR flare \citep{Malyali2021,Onori2022,Short2023}. Only one of our sources (WTP\,18aampwj) shows the formation of coronal lines nearly 5 years after the initial transient, but only [Fe VII] lines (IP $\approx$ 99 eV) are present in the spectrum. The remaining sources in our sample do not show transient coronal lines, although we do see transient forbidden lines commonly associated with accretion-driven ionization, including narrow [O III] and [O I] in WTP\,17aamoxe and WTP\,17aamzew (see Figure \ref{fig:wtp17s}). Potentially the lack of evidence for coronal lines in our sample could be due to the timescales probed with optical spectra; in particular, many of the early disruptions in our sample only have follow-up nearly 8-9 years after the initial disruption. This link between the IR and coronal emission lines suggests a common origin -- TDEs in gas- and dust-rich environments.

A rather surprisingly large number of our IR TDEs show evidence for X-ray emission from eROSITA. Of the eight sources in the German half of the sky, three of them show late-time ($\gtrsim$ 2 yrs), low-luminosity ($L_X \approx 10^{41-42}$~erg~s$^{-1}$) X-ray emission, with a variety of spectral shapes. Two sources show a relatively soft X-ray spectrum (WTP\,17aamoxe and WTP\,17aamzew; $\Gamma \approx 3-4$ or $kT_\mathrm{bb} \approx 0.1-0.15$ keV), and one source shows a relatively hard X-ray spectrum (WTP\,15abymdq; $\Gamma \approx 1.7$). While soft X-ray emission associated with the inner accretion disk is often what is dominant in most X-ray-bright TDEs, harder X-ray emission, often associated with the formation of an X-ray corona, is not unheard of, especially at late times \citep[e.g.][]{Saxton2017,Kara2018,Chakraborty2021,Yao2022}. In our sources, the presence of soft X-ray emission is somewhat puzzling, as obscuration by gas along our line of sight should primarily attenuate the soft X-ray band. This could suggest that the nucleus has been unveiled at late times due to a patchy or clumpy obscuring medium. Such a picture is also consistent with the rare occurrence of optical counterparts in a few of our sources. Interestingly, the three sources in our sample with eROSITA detections are X-ray bright for three or more eRASS epochs and all are detected more than five years since the initial disruption, thus motivating further late-time X-ray follow-up of optically-selected TDEs \citep[see e.g.][for recent work on this topic]{Jonker2020}.

It is important to note that due to the lack of an sensitive, all-sky X-ray survey until eROSITA came online in 2019 and the rarity of optical flares in the WISE TDEs, we lack sufficient X-ray coverage during the peak of most of the WISE flares. Hence, while eROSITA allows us to study the late-time X-ray properties of our sample, we cannot claim an absence of X-ray counterparts in other sources which are undetected in late-time X-ray observations. Future work, either with WISE-detected IR flares from 2019 onward (i.e. after eROSITA came online) or with more rapid detection of IR flares with Roman for example, will be crucial to probing the prompt X-ray emission in IR-selected TDEs.

\subsection{No Evidence for Neutrino Counterparts}

Recent advances in the detection of high-energy astrophysical neutrinos with the IceCube Observatory have opened a new window into the multi-messenger sky. In recent years, a few TDE candidates have been claimed to have a spatially and temporally coincident high-energy neutrino \citep[e.g.][]{Stein2021,Jiang2023}, leading to the suggestion that TDEs could contribute a sizeable fraction of the cosmic high-energy neutrino flux \citep{vanVelzen2021a}. Interestingly, the detected TDE/neutrino sources all showed strong IR dust echoes, suggesting a potential link between the accretion flare and surrounding medium, although the underlying mechanisms are still poorly understood \citep[e.g.][]{Winter2023}. Our sample of IR-selected TDEs is thus a natural place to search for coincident neutrinos. We cross-matched between the gold WISE TDEs and the gold candidate neutrinos ($\gtrsim$ 50\% probability of being astrophysical) from the most recent IceCube catalog \citep[IceCat-1;][]{Abbasi2023}. We found no spatially and temporally coincident neutrinos using the 90\% uncertainty regions given by IceCube. However, given the low number of TDEs in our gold sample and the relatively low number of gold neutrinos, we cannot statistically rule out that the IR-emitting TDEs are sources of high-energy neutrinos. We plan to explore  the statistical significance and theoretical interpretation of these findings further in future work.

\section{Summary \& Conclusions} \label{sec:summary}

We have presented a new sample of IR-selected TDEs, using difference photometry from the WISE satellite to identify transient mid-IR emission in the nuclei of spectroscopically confirmed galaxies within roughly 200 Mpc ($z \lesssim 0.05$). We identified 18 TDE candidates, which showed: (1) peak IR emission above $L_\mathrm{W2} \gtrsim 10^{42}$~erg~s$^{-1}$, (2) WISE light curves with a fast rise, followed by a slow, monotonic decline, (3) no significant prior IR variability, and (4) no AGN-like WISE colors or previous ROSAT X-ray detection. To further remove potential AGN contaminants, we excluded any source with narrow line ratios in optical spectroscopy indicative of AGN ionization processes. After these cuts to remove sample contaminants, our final gold sample contains 12 high-confidence TDEs. 

This work represents the cleanest sample of IR-selected TDEs to date. As such, it presents new insights into TDE demographics in the local universe, and allows us to probe the properties of a previously overlooked class of TDEs. Our main findings with this sample of 12 IR-selected TDEs are summarized below: 
\begin{itemize}
    \item This new sample of IR-selected TDEs shows very few optical counterparts, with only one of our 12 TDEs showing a weak optical flare. The lack of an optical counterpart, coupled with such a strong IR flare, suggests that these sources are a new population of obscured TDEs that were missed by other surveys. 
    \item We modeled the IR flares with a spherical shell model for the dust to derive properties such as the dust radius and disruption time. The average dust radius is $R_\mathrm{dust} = 0.23$~pc, with a range from 0.05-0.46~pc among all sources. The small range of dust radii is a natural consequence of the fast rise, slow decline light curve shape required in our selection criteria. Together with blackbody modeling of the WISE data, we constrain the covering factor in these sources to be $f_c \gtrsim 10\%$, which is in stark contrast to the roughly 1\% covering factor seen in optical TDEs. 
    \item Four sources showed transient accretion-driven emission lines, including the appearance and disappearance of a broad H$\alpha$ line in one source, the appearance of a broad He I time in another, and the appearance of coronal lines in one source and forbidden lines like [O III] and [O I] in others.
    \item A surprisingly high fraction of sources (3/8) in the German half of the eROSITA sky were detected in the X-ray band in numerous epochs. As with other late-time X-ray constraints of TDEs, these sources showed a variety of X-ray spectral shapes, ranging from rather hard ($\Gamma \approx 1.7$) to rather soft ($\Gamma \approx 4$).
    \item The TDEs all radiate between $E_\mathrm{rad,\, bb} \approx 4 \times 10^{50}$-$9 \times 10^{51}$ erg in the IR alone, which is pushing the limits on the IR emission from SNe. When combined with evidence for transient line emission and X-ray emission, this suggests that these events are driven by accretion power.
    \item We derive a lower limit on the rate of these IR TDEs, finding a rate of $2.0 \pm 0.3 \times 10^{-5}$ galaxy$^{-1}$ year$^{-1}$ and a corresponding volumetric rate of $1.3 \pm 0.2 \times 10^{-7}$ Mpc$^{-3}$ year$^{-1}$. This rate is comparable to both the optical and X-ray TDE rate in the local universe, suggesting that dust obscured TDEs are about as common as optical and X-ray TDEs. Together, these three populations are inching closer to the theoretically predicted rate of $10^{-4}$ galaxy$^{-1}$ year$^{-1}$.
    \item Our host galaxy analysis reveals that the IR TDEs do not show the same green valley or post-starburst overdensity that is often seen in optical and X-ray TDEs. This highlights potential differences in host galaxy selection effects of TDE searches at different wavelengths.
\end{itemize}


\smallskip

\noindent  We thank the anonymous referee for their insightful comments. We thank Brad Cenko and the Swift team for approving the numerous ToO requests associated with this sample. We thank Mansi Kasliwal, Josh Bloom, and the SkyPortal team for assistance with using the data platform for this project. We thank Yuhan Yao for discussions about host galaxy SED fitting, Morgan MacLeod for discussions on partial covering of the accretion regions, and Collin Lewin for discussions on transfer function modeling. K. D. was supported by NASA through the NASA Hubble Fellowship grant \#HST-HF2-51477.001 awarded by the Space Telescope Science Institute, which is operated by the Association of Universities for Research in Astronomy, Inc., for NASA, under contract NAS5-26555. I. A. acknowledges support from the European Research Council (ERC) under the European Union’s Horizon 2020 research and innovation program (grant agreement number 852097), from the Israel Science Foundation (grant number 2752/19), and from the United States - Israel Binational Science Foundation (BSF). A. M. acknowledges support by DLR under the grant 50 QR 2110 (XMM\_NuTra). \\

\noindent This publication makes use of data products from the Wide-field Infrared Survey Explorer and NEOWISE, which are joint projects of the University of California, Los Angeles, and the Jet Propulsion Laboratory/California Institute of Technology, funded by the National Aeronautics and Space Administration. This paper includes data gathered with the 6.5 meter Magellan Telescopes located at Las Campanas Observatory, Chile. This work makes use of observations from the Las Cumbres Observatory global telescope network. This work is based on observations obtained at the Southern Astrophysical Research (SOAR) telescope, which is a joint project of the Minist\'{e}rio da Ci\^{e}ncia, Tecnologia e Inova\c{c}\~{o}es (MCTI/LNA) do Brasil, the US National Science Foundation’s NOIRLab, the University of North Carolina at Chapel Hill (UNC), and Michigan State University (MSU). Visiting Astronomer at the Infrared Telescope Facility, which is operated by the University of Hawaii under contract 80HQTR19D0030 with the National Aeronautics and Space Administration. This paper employs a list of Chandra datasets, obtained by the Chandra X-ray Observatory, contained in \dataset[DOI: 10.25574]{https://doi.org/10.25574/cdc.183}. \\

\noindent This work is based on data from eROSITA, the soft X-ray instrument aboard SRG, a joint Russian-German science mission supported by the Russian Space Agency (Roskosmos), in the interests of the Russian Academy of Sciences represented by its Space Research Institute (IKI), and the Deutsches Zentrum für Luftund Raumfahrt (DLR). The SRG spacecraft was built by Lavochkin Association (NPOL) and its subcontractors and is operated by NPOL with support from the Max Planck Institute for Extraterrestrial Physics (MPE). The development and construction of the eROSITA X-ray instrument were led by MPE, with contributions from the Dr. Karl Remeis Observatory Bamberg \& ECAP (FAU Erlangen-Nuernberg), the University of Hamburg Observatory, the Leibniz Institute for Astrophysics Potsdam (AIP), and the Institute for Astronomy and Astrophysics of the University of Tübingen, with the support of DLR and the Max Planck Society. The Argelander Institute for Astronomy of the University of Bonn and the Ludwig Maximilians Universität Munich also participated in the science preparation for eROSITA.


\facilities{WISE/NEOWISE, Swift, SRG/eROSITA, Chandra, XMM-Newton, Magellan (Clay/LDSS3, Baade/FIRE), SOAR/GHTS, IRTF/SpeX, Las Cumbres/FLOYDS, Gemini-S/GMOS, ASAS-SN, ATLAS, ZTF}

\software{XSPEC (v12.12.0; \citealt{Arnaud1996}),
\texttt{pypeit} \citep{Prochaska2020},
FSPS \citep{Conroy2009,Conroy2010},
\texttt{prospector} \citep{Johnson2021},
\texttt{emcee} \citep{Foreman-Mackey2013},
\texttt{astropy} \citep{AstropyCollaboration2022},
eSASS}

\begin{appendix} 

\begin{deluxetable*}{c c c c c c c c}
	\caption{Additional Optical and Near-IR Spectroscopic Follow-Up of IR TDE Candidates} \label{tab:spec}
    \tablehead{\colhead{ID} & \colhead{WTP Name} & \colhead{Date} & \colhead{Telescope} & \colhead{Instrument} & \colhead{Wavelength Range (\AA)} & \colhead{Slit Width (\arcsec)} & \colhead{Exp. Time (s)}}
	\startdata
	1 & WTP\,14abnpgk & 2023-05-27 & Magellan Clay & LDSS3 & 4250-10000 & 0.75 & 1800 \\
    & & 2023-06-24 & IRTF & SpeX & 8000-25000 & 0.5 & 2160 \\
    2 & WTP\,14acnjbu & 2023-05-27 & Magellan Clay & LDSS3 & 4250-10000 & 0.75 & 1800 \\
    & & 2023-06-24 & IRTF & SpeX & 8000-25000 & 0.5 & 2880 \\
    3 & WTP\,14adbjsh & 2022-07-04 & Magellan Baade & FIRE (Echelle) & 8000-25000 & 0.6 & 1800 \\
    & & 2022-11-11 & SOAR & GHTS (red) & 6300-8930 &  1.0 & 1200 \\
    & & 2022-11-14 & SOAR & GHTS (blue) & 3500-7000 &  1.0 & 600 \\
    4 & WTP\,14adbwvs & 2022-05-09 & FTN & FLOYDS & 3500-10000 & 2.0 & 1800 \\
    & & 2023-08-25 & IRTF & SpeX & 8000-25000 & 0.5 & 2880 \\
    5 & WTP\,14adeqka & 2022-04-06 & FTN & FLOYDS & 3500-10000 & 2.0 & 1800 \\
    & & 2023-06-24 & IRTF & SpeX & 8000-25000 & 0.5 & 2160 \\
    6 & WTP\,15abymdq & 2023-02-26 & Magellan Clay & LDSS3 & 4250-10000 & 0.75 & 1800 \\
    7 & WTP\,15acbgpn & 2023-04-06 & FTN & FLOYDS & 3500-10000 & 2.0 & 1800 \\
    8 & WTP\,15acbuuv & 2023-02-26 & Magellan Clay & LDSS3 & 4250-10000 & 0.75 & 1800 \\
    9 & WTP\,16aaqrcr & 2023-05-27 & Magellan Clay & LDSS3 & 4250-10000 & 0.75 & 1800 \\
    10 & WTP\,16aatsnw & 2023-05-27 & Magellan Clay & LDSS3 & 4250-10000 & 0.75 & 1800 \\
    & & 2023-08-25 & IRTF & SpeX & 8000-25000 & 0.5 & 2880 \\
    11 & WTP\,17aaldjb & 2023-05-27 & Magellan Clay & LDSS3 & 4250-10000 & 0.75 & 1800 \\
    & & 2023-06-24 & IRTF & SpeX & 8000-25000 & 0.5 & 2880 \\
    12 & WTP\,17aalzpx & 2023-02-26 & Magellan Clay & LDSS3 & 4250-10000 & 0.75 & 1800 \\
    13 & WTP\,17aamoxe & 2023-05-27 & Magellan Clay & LDSS3 & 4250-10000 & 0.75 & 1800 \\
    &  & 2023-05-27 & Gemini-S & GMOS & 3600-10300 &  1.0 & 1050 \\
    14 & WTP\,17aamzew & 2023-02-26 & Magellan Clay & LDSS3 & 4250-10000 & 0.75 & 1800 \\
    & & 2023-03-21 & IRTF & SpeX & 8000-25000 & 0.5 & 1440 \\
    15 & WTP\,17aanbso & 2023-04-06 & FTN & FLOYDS & 3500-10000 & 2.0 & 1800 \\
    16 & WTP\,18aajkmk & 2023-02-26 & Magellan Clay & LDSS3 & 4250-10000 & 0.75 & 1800 \\
    &  & 2023-01-17 & Magellan Baade & FIRE & 8000-25000 & 0.6 & 300 \\
    18 & WTP\,18aampwj & 2023-07-25 & SOAR & GHTS (blue) & 3500-7000 &  1.0 & 600 \\
    & & 2023-08-25 & IRTF & SpeX & 8000-25000 & 0.5 & 2880 \\
    \enddata
\end{deluxetable*}

\section{Spectroscopic Follow-Up} \label{sec:appendix_specfollowup}

\begin{figure*}[t!]
    \centering
    \includegraphics[width=\textwidth]{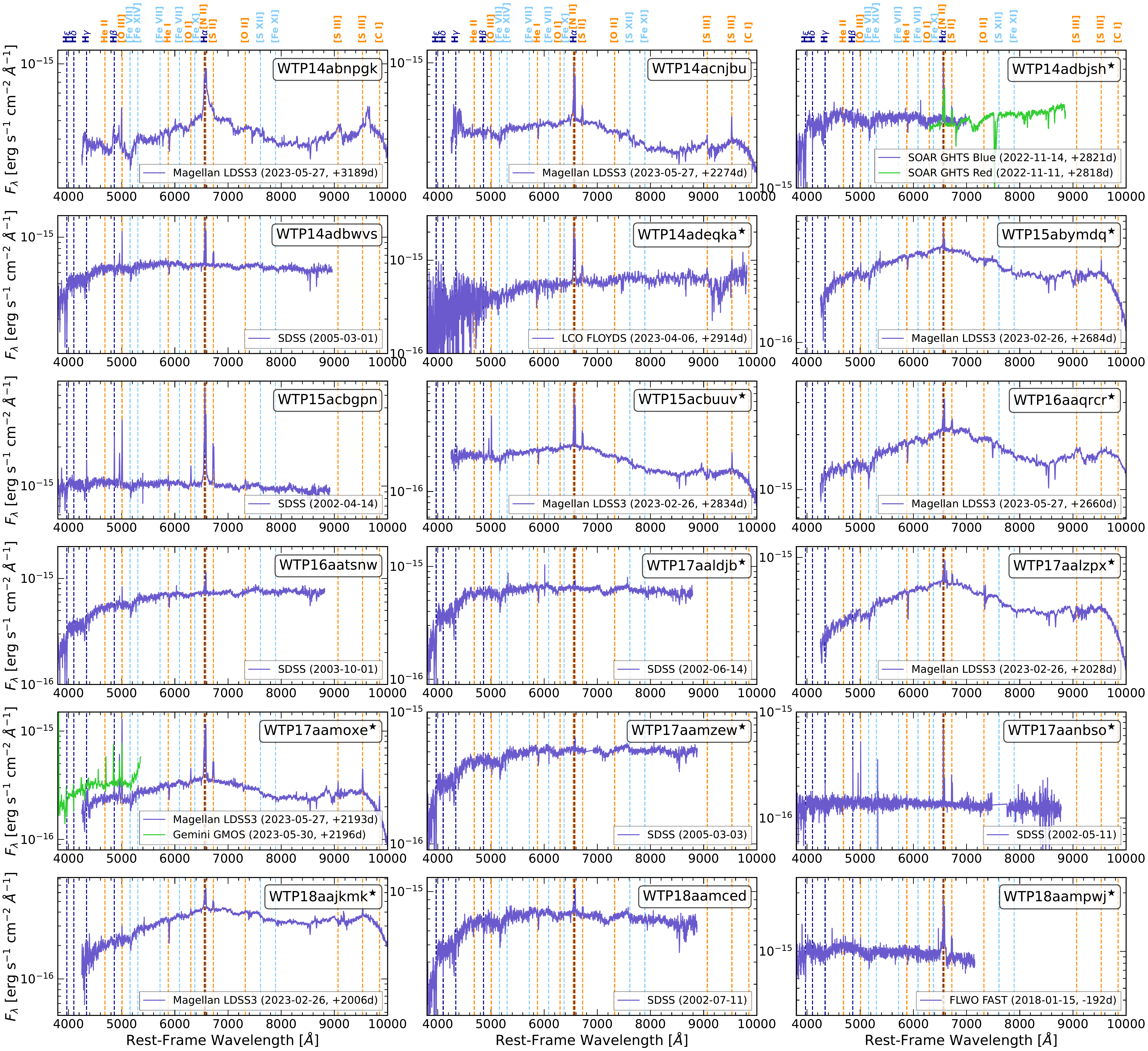}
    \caption{Optical spectroscopy of each of the TDE candidates in our sample. For those sources with archival SDSS spectra, we show those here, instead of post-flare observations. The vertical lines show important spectral lines, with Balmer lines shown in dark purple, coronal lines in light blue, and other important lines of interest in orange. All gold sample TDEs are marked with a $^\bigstar$. Detailed comments on the AGN-like nature of the silver sample, evidence of spectral evolution, and existence of broad emission lines are given in the individual source section (Section \ref{sec:appendeix_indiv}). }
    \label{fig:opt_spec}
\end{figure*}

\begin{figure*}[t!]
    \centering
    \includegraphics[width=\textwidth]{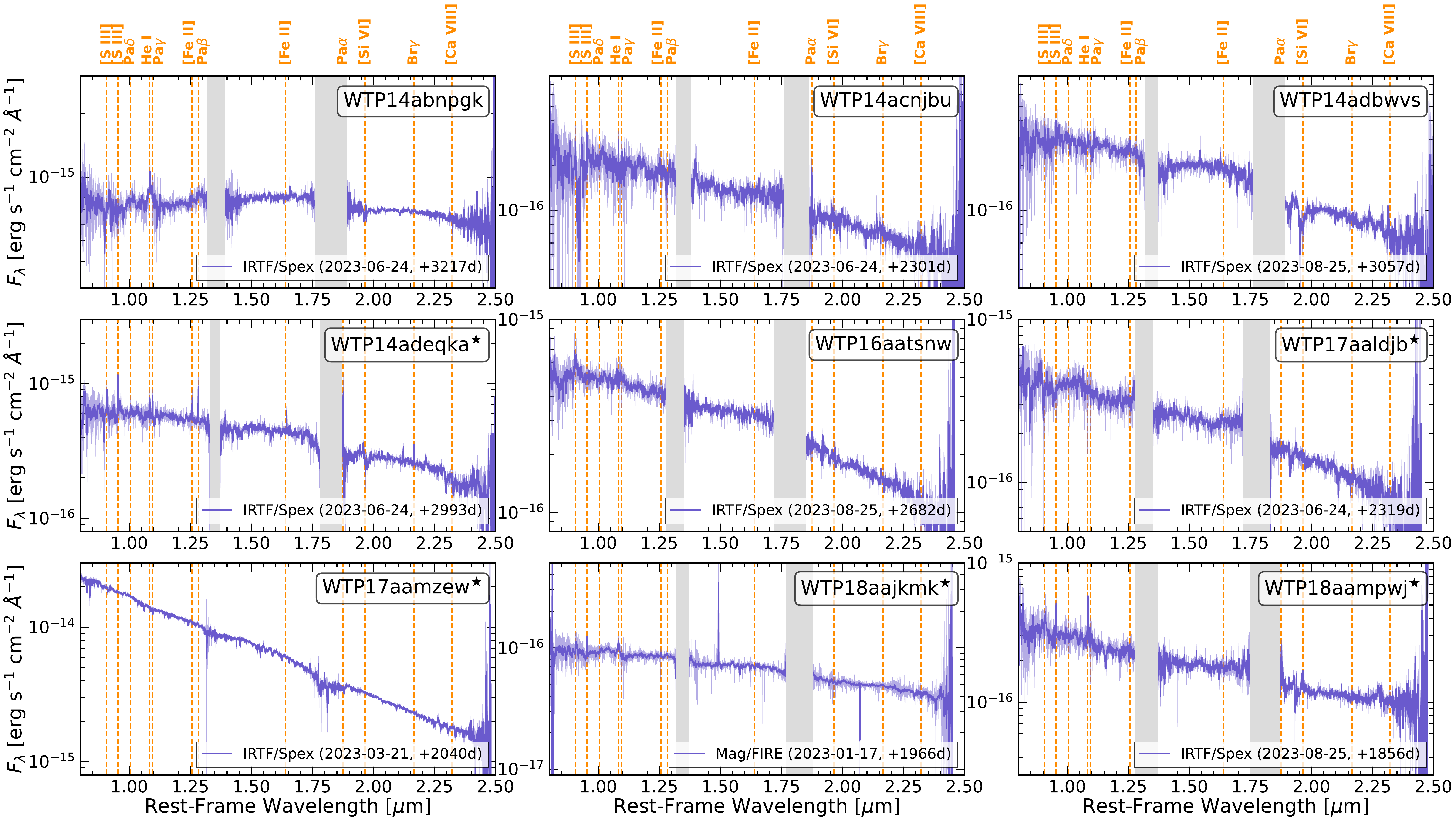}
    \caption{Post-flare near-IR spectroscopy from either Magellan/FIRE or IRTF/SpeX of nine of the TDE candidates in our sample with near-IR spectroscopy to date. The light color spectrum is the raw spectrum, and the dark, thicker line shows the spectrum binned to a resolution of 10\AA. The vertical orange dashed lines show important spectral lines in the near-IR. Regions of strong telluric absorption are shown as grey shaded regions. All gold sample TDEs are marked with a $^\bigstar$. Only WTP\,14abnpgk (silver sample; likely AGN) and WTP\,18aajkmk (gold sample; likely TDE) show any broad lines (He I, 1.083 $\mu$m).}
    \label{fig:nir_spec}
\end{figure*}

In Table \ref{tab:spec}, we list the details of the optical and near-IR spectroscopic follow-up observations we performed of the sources in our sample. We also present an optical spectrum for each source in Figure \ref{fig:opt_spec}. For any source where an archival SDSS spectrum exists, we present this spectrum as it is best for ruling out AGN contamination. Additionally, for the nine sources with near-IR spectroscopic follow-up, we present these spectra in Figure \ref{fig:nir_spec}.

\section{Host Galaxy SED Fits} \label{sec:appendix_sed}

\begin{figure*}[t!]
    \centering
    \includegraphics[width=\textwidth]{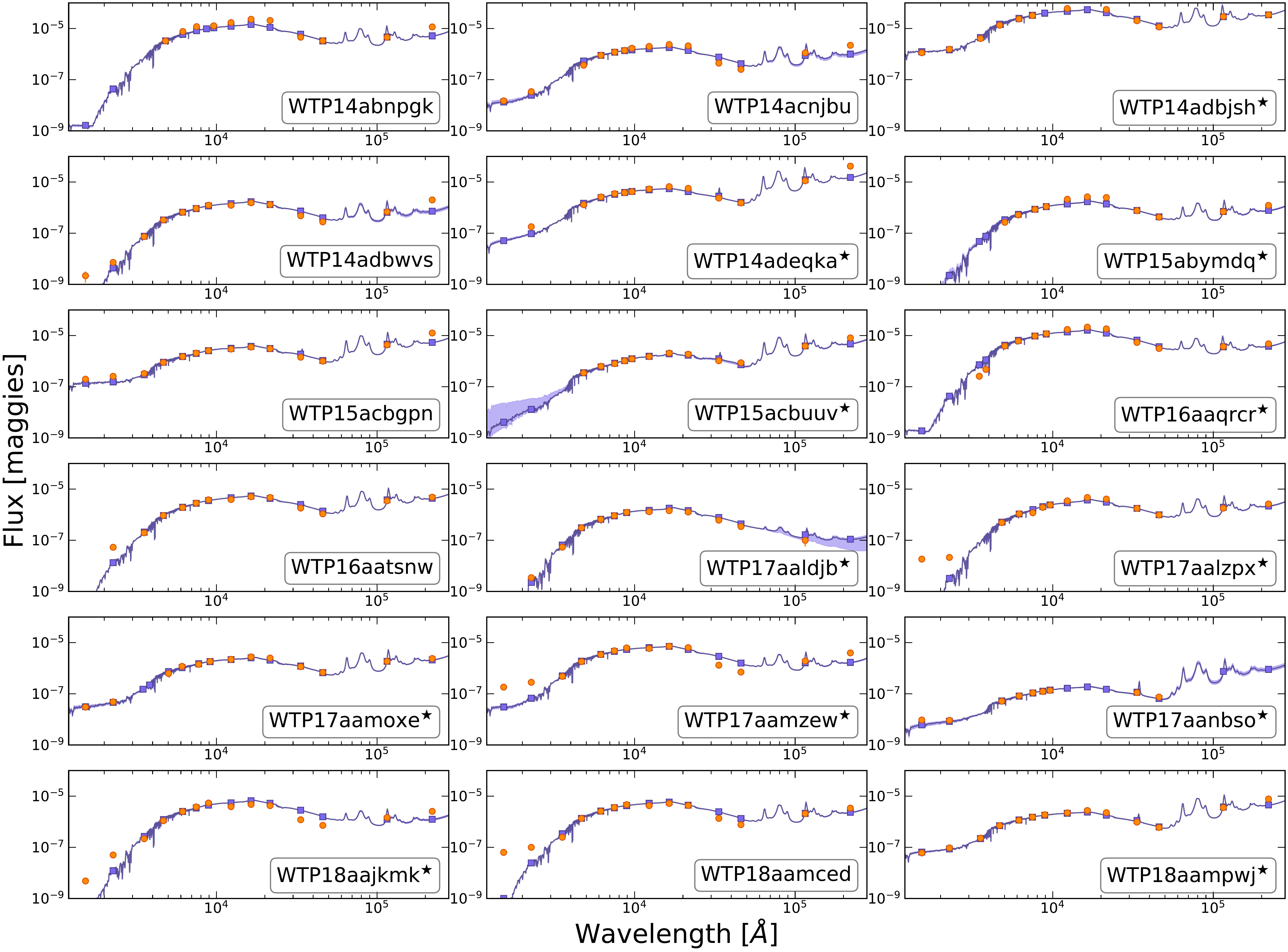}
    \caption{Host galaxy photometry and SED fits with \texttt{prospector}. The orange points show the observed photometry, which has been corrected for Galactic extinction. The purple points show the best-fit model photometry, with the modeling details given in Section \ref{subsec:seds}. The shaded purple regions show the uncertainty regions from the MCMC modeling. All gold sample TDEs are marked with a $^\bigstar$.}
    \label{fig:SEDs}
\end{figure*}

In Figure \ref{fig:SEDs}, we show the resulting SED fits for each source from Section \ref{subsec:seds}. 

\section{Comments on Individual Sources} \label{sec:appendeix_indiv}

In this Section, we provide individual details on each source, as well as our reasoning for classifying it in either the gold or silver sample. 

\subsection{WTP\,14abnpgk}

\begin{figure*}[t!]
    \centering
    \includegraphics[width=\textwidth]{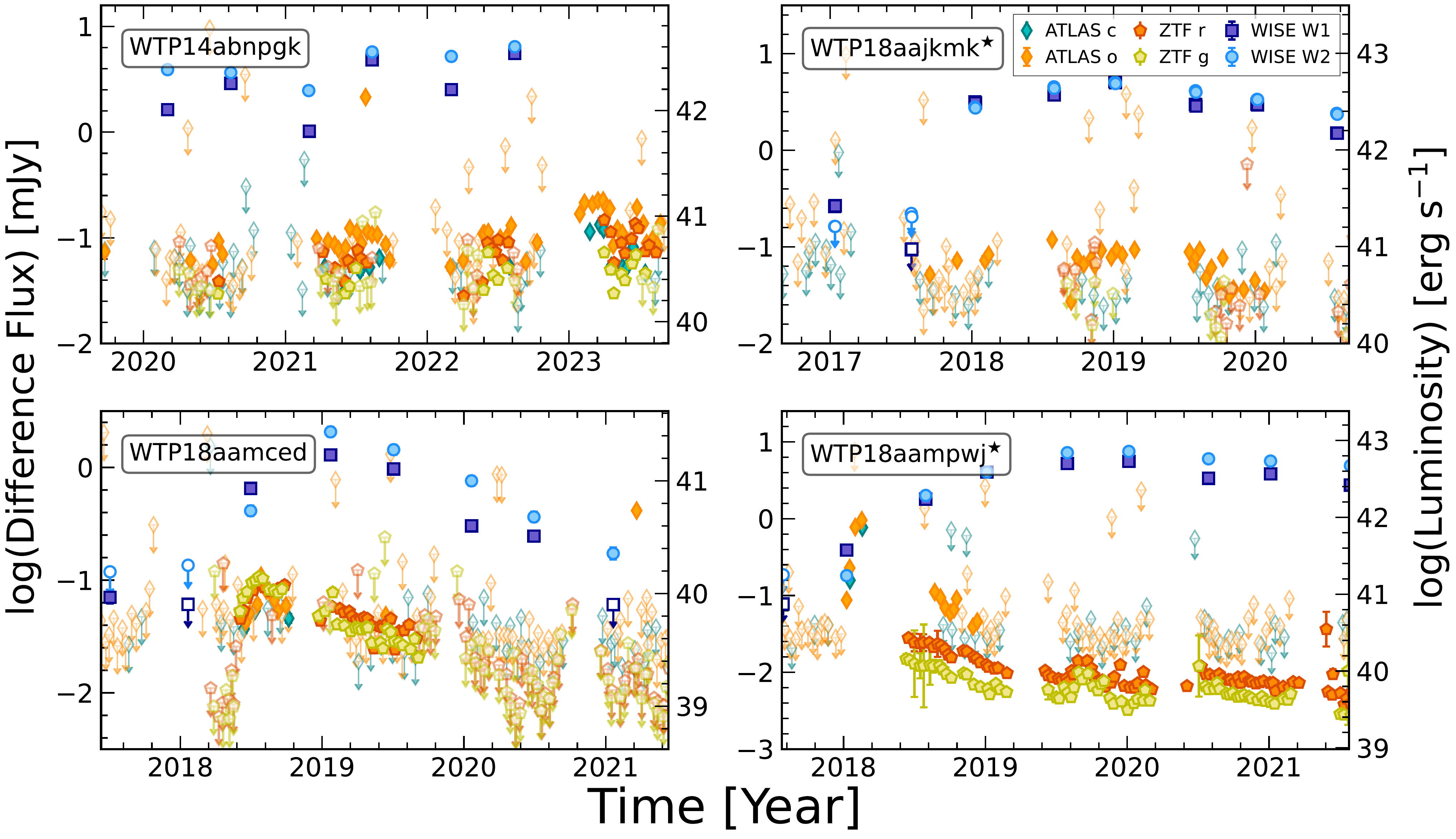}
    \caption{Zoom in on the optical (ATLAS, ZTF) and IR (WISE) light curves for the four sources that show any hint of an optical transient. The optical data from ATLAS (cyan and orange diamonds) and ZTF (yellow and red pentagons) have been binned on 10 day timescales using an inverse variance weighted average with 3$\sigma$ clipping for outlier rejection. Each panel shows a 4 year duration, zoomed in around the time of the optical/IR flare. Gold sample sources are marked with a $^\bigstar$. The left axis shows the log of the difference flux in mJy, and the right axis shows the corresponding monochromatic W2 luminosity in~erg~s$^{-1}$. The optical flares in these sources are more than an order of magnitude fainter than the IR flare, which is unlike most optically-selected TDEs. \textit{Top left:} Light curves for WTP\,14abnpgk, focused on the secondary rise in the WISE light curve that started around 2021-2022. ZTF and ATLAS both show significant detections around this time frame, although the light curves are more structured than is typically seen in optical TDE light curves. \textit{Top right:} Light curves for WTP\,18aajkmk, which did not have a previously-detected optical transient. ATLAS shows a potential flare that appears to be contemporaneous with the IR flare, but no flare is detected with ZTF. \textit{Bottom left:} Light curves for WTP\,18aamced, which was extensively studied in \cite{Frederick2019} and \cite{Huang2023}. The optical transient (AT\,2018dyk) was detected in both ZTF and ATLAS, with a light curve that resembles typical optically-selected TDEs. \textit{Bottom right:} Light curves for WTP\,18aampwj, which was detected as an optical transient (AT\,2018gn) in 2018 by ASAS-SN. The optical transient was detected in ASAS-SN, ATLAS, and ZTF. However, the transient first went off when ZTF was taking reference images, so the exact flux scaling of the ZTF data shown here is not correct.}
    \label{fig:optcounterparts}
\end{figure*}

WTP\,14abnpgk showed a strong and relatively smooth IR flare, but has recently shown a secondary rise in both W1 and W2 in the last three WISE epochs. Interestingly, both ZTF and ATLAS detected an  optical transient in this source in 2022, around the same time as the second rise is seen in WISE. The optical transients was reported to TNS as a nuclear transient (AT\,2022mim\footnote{\url{https://www.wis-tns.org/object/2022mim}}). The ATLAS and ZTF light curves, shown in Figure \ref{fig:optcounterparts}, do not look like typical optical TDE light curves and instead resemble the more stochastic variability often seen in AGN. In addition, an optical spectrum taken with the LDSS3 instrument on the Magellan Clay telescope revealed a broad H$\alpha$ line with a significant red wing and diminished blue wing, as well as a similar line structure in the H$\beta$ profile. Additional near-IR follow-up showed a broad He I line at 1.083 $\mu$m, although there is potential blending with the Pa$\alpha$ line. These broad lines are also suggestive that the source is an AGN, although it is also possible that this is a partial TDE. A double dust echo has not been reported in repeated TDEs previously, but repeated optical and X-ray flares have been both been used to claim partial TDEs \citep[e.g.][]{Payne2021,Liu2023,Wevers2023}. As there are no public archival optical spectra of this source, we cannot probe the evolution of the broad emission lines. Hence, further monitoring is necessary to disentangle the exact nature of this source. To be conservative, we place this source in our silver sample, due to the potential indications of AGN activity.


\subsection{WTP\,14acnjbu}

WTP\,14acnjbu has the most peculiar WISE light curve of any source in our sample, showing a slight rise before the TDE-like flare. With the LDSS3 spectrum from 2023, we find that the source sits in either the star-forming or composite region of BPT parameter space. However, to be conservative, we place this source in our silver sample, as the double peaked light curve could be indicative of veiled AGN activity.

\subsection{WTP\,14adbjsh}

WTP\,14adbjsh, the brightest and closest source in our sample, was already presented in detail in \cite{Panagiotou2023}. In short, we find no optical/near-IR spectral lines associated with accretion, which suggests that the IR flare is not driven by an AGN. Late-time X-ray observations with Swift and NuSTAR revealed no significant X-ray emission. We place WTP\,14adbjsh in our gold sample. WTP\,14adbjsh occurred in a star-forming galaxy and is one of the closest TDEs discovered to date, despite being entirely missed by optical and X-ray searches.

\subsection{WTP\,14adbwvs}

WTP\,14adbwvs is classified as an AGN, based on its location in the BPT diagrams from archival SDSS spectroscopy. We obtained a recent spectrum with FLOYDS and found no evidence for any spectral changes. The source does not have strong X-ray emission, nor is its WISE color indicative of AGN activity. However, out of an abundance of caution, we place this source in our silver sample.

\subsection{WTP\,14adeqka}

WTP\,14adeqka is one of the strongest and smoothest flares in our sample, reaching a peak luminosity around $10^{43}$~erg~s$^{-1}$. The host galaxy of WTP\,14adeqka shows no evidence for significant AGN-like ionization (e.g. [O III], [O I], [S II]) when we observed the source in 2023 with FLOYDS. The spectral resolution of FLOYDS leaves us unable to resolve the H$\alpha$ + [N II] complex, but no broad lines are observed. We also find no evidence for broad lines in a near-IR spectrum taken with IRTF/SpeX, but this spectrum does show evidence for some forbidden narrow lines (e.g. [S III] 9530\AA, [Fe II] 1.257$\mu$m, 1.64$\mu$m). However, other sources have shown late-time evidence for forbidden narrow lines associated with the TDE (i.e. no such lines existed before the IR flare). Hence, the existence of forbidden lines alone does not suggest AGN-drive behavior. Given that there are no other indicators of an underlying AGN, we place WTP\,14adeqka in our gold sample.

\subsection{WTP\,15abymdq}

WTP\,15abymdq shows no strong evidence for an AGN in the optical spectrum we obtained with LDSS3 in 2023, sitting in the star-forming or composite region of parameter space for various BPT diagrams. This source was one of the three that we detected with eROSITA, and the spectrum was relatively hard, showing a power-law X-ray spectrum with $\Gamma \approx 1.7$. There is evidence for neutral absorption with a variable column density on the order of $10^{21}$ cm$^{-2}$ (see Section \ref{subsec:xray} for more details). Although this is a relatively standard AGN-like X-ray spectrum, the X-ray spectra of TDEs are not uniform enough for us to use this to rule out this source as a TDE. Hence, we place WTP\,15abymdq in our gold sample, given that there are no other indications of underlying AGN activity in this source.

\subsection{WTP\,15acbgpn}

WTP\,15acbgpn is a clear AGN with a broad H$\alpha$ line apparent in the SDSS spectrum from 2002. It was also detected with Chandra in 2010, with an X-ray flux of $F_{0.5-7\,\mathrm{keV}} \approx 8 \times 10^{-13}$ erg cm$^{-2}$ s$^{-1}$. Thus, given the clear AGN nature, we place this source in our silver sample. We note that we did observe this source with FLOYDS in 2023 and found that the spectrum was consistent with the SDSS spectrum, albeit with lower spectral resolution. 

\subsection{WTP\,15acbuuv}

WTP\,15acbuuv sits in the star-forming or composite region of BPT parameter space, although there is significant [O III] emission in this spectrum. A past spectrum from 6dF \citep{Jones2004,Jones2009} shows little evolution from the LDSS3 spectrum we took in 2023. This source was also one of the eROSITA sources, and showed no evidence for any X-ray emission during any of the eRASS epochs. Given the lack of significant evidence for AGN behavior, we place this source in our gold sample.

\subsection{WTP\,16aaqrcr}

WTP\,16aaqrcr shows no clear indications of any previous AGN activity. An optical spectrum from LDSS3 in 2023 shows no evidence of [O III] emission, consistent with star-forming ionization. The source was also never detected in the eROSITA survey. Hence, we place WTP\,16aaqrcr in our gold sample.

\subsection{WTP\,16aatsnw}

WTP\,16aatsnw is another potential AGN/LINER interloper in our sample, based on its location in the BPT diagrams from archival SDSS spectroscopy. We obtained a spectrum with LDSS3 in 2023 and found no evidence for significant spectral changes. The source does not have strong X-ray emission, nor is its WISE color indicative of AGN activity. However, out of an abundance of caution, we place this source in our silver sample, given the pre-flare AGN-like ionization levels.

\subsection{WTP\,17aaldjb}

WTP\,17aaldjb has an archival SDSS spectrum which shows no evidence of AGN activity, with only very weak emission lines visible in the spectrum. Likewise, our LDSS3 spectrum from 2023 was very similar, showing little evolution. We also obtained a near-IR spectrum of this source with IRTF/SpeX in 2023 and found similarly negligible emission features, strengthening our confidence that this is not an veiled AGN. The source was also not detected with eROSITA. Hence, we place this source in our gold sample.

\subsection{WTP\,17aalzpx}

WTP\,17aalzpx showed no strong [O III] emission in the LDSS3 spectrum from 2023, indicating likely star-forming ionization. The source was also not detected with eROSITA. Hence, this source does not show any strong evidence for AGN activity, and thus we place it in our gold sample.

\subsection{WTP\,17aamoxe}

WTP\,17aamoxe showed very strong, AGN-like narrow emission lines in the optical spectrum with obtained with LDSS3 in 2023. However, the archival spectrum from 2004 with the 6dF survery \citep{Jones2004,Jones2009} shows no significant [O III], [O I], or [S II] emission, indicating that there has likely been a significant change to the spectral lines in this source (see Section \ref{subsec:transient_lines}). WTP\,17aamoxe is also one of the three sources that we detected with eROSITA, showing a relatively soft X-ray spectrum ($\Gamma \approx 4$) in the three eRASS epochs in which it was detected (see Section \ref{subsec:xray}). Such a soft X-ray spectrum is highly unusual for AGN, but is often seen in TDEs. Hence, we place this source in our gold sample, but note that further follow-up to monitor the X-ray and optical spectral changes is necessary to rule out a turn-on AGN.

\subsection{WTP\,17aamzew}

WTP\,17aamzew is another one of our three detected eROSITA sources, again showing a soft X-ray spectrum ($\Gamma \approx 4$). This source has an archival SDSS spectrum from 2005, which shows no evidence for AGN-driven line emission. However, our LDSS3 spectrum from 2023 shows significant spectral changes, including the formation of both [O III] and [O I], which are often associated with accretion. We suspect that these narrow lines were induced by the TDE. Hence, we place WTP\,17aamzew in our gold sample.

\subsection{WTP\,17aanbso}

WTP\,17aanbso shows strong emission lines in its archival SDSS spectrum, but the line ratios suggest a purely star-forming nature. We obtained an FLOYDS spectrum in 2023, but found no significant evidence for spectral changes. Hence, as there is no evidence for underlying AGN activity, we place WTP\,17aanbso in our gold sample.

\subsection{WTP\,18aajkmk}

WTP\,18aajkmk shows no clear evidence for ongoing AGN activity, falling primarily in the star-forming or composite region of BPT parameter space. The [O I] line from our 2023 LDSS3 spectrum does show LINER ionization levels, but there are other sources with significant narrow line changes and we have no pre-outburst optical spectrum. Hence, we cannot rule out that this line is driven by the TDE. Additionally, we do see broad He I line at 1.083 $\mu$m in a FIRE near-IR spectrum taken in 2023, which we believe is likely associated with the TDE. Interestingly, this source is clearly an extremely disturbed system (see Figure \ref{fig:cutouts}), thereby potentially suggesting a merger-driven accretion episode. {WTP\,18aajkmk also shows potential evidence for an optical transient in the ATLAS $o$-band data (see top left of Figure \ref{fig:optcounterparts}). No transient is detected in the $c$-band data, which could be due to the reddening from a dusty environment. ZTF does not detect a significant optical transient in this system, although this could be a result of a transient occurring around the time of reference images being taken.} Given the lack of any evidence for AGN activity, we classify this source as a likely TDE, and place it in our gold sample.

\subsection{WTP\,18aamced}

WTP\,18aamced is one of the few sources in our sample with a clear optical counterpart, first detected by the the ZTF team\footnote{\url{https://www.wis-tns.org/object/2018dyk}}. We show the ZTF and ATLAS light curves for this source with its WISE light curve in the bottom left panel of Figure \ref{fig:optcounterparts}. The ZTF source was relatively faint compared to other TDEs, and extensively optical and UV spectroscopic follow-up showed that this source transitioned from a LINER (based on the archival SDSS spectrum) to a narrow-line Seyfert 1 with broad Balmer lines and strong Fe and He coronal lines \citep{Frederick2019}. Like many narrow-line Seyfert 1s, the source shows a relatively soft X-ray spectrum ($\Gamma \approx 2.8-3$), but is still harder than the two other soft sources in our sample \citep{Frederick2019}. It has recently been argued, however, that this source is a TDE in a LINER galaxy with a dusty nuclear environment \citep{Huang2023}. For detailed studies of this source, we refer the reader to \cite{Frederick2019} and \cite{Huang2023}. As the source shows clear LINER ionization levels before the flare, we place it in our silver sample.

\subsection{WTP\,18aampwj}

WTP\,18aampwj was first discovered by the ASAS-SN team\footnote{\url{https://www.wis-tns.org/object/2018gn}} as a candidate Type II SN \citep{Falco2018}, after an optical spectrum revealed a broad H$\alpha$ line (see Figure \ref{fig:broad_lines}). The optical transient was also clearly detected in the deeper ATLAS and ZTF surveys, although the optical flare occurred when the ZTF reference images were being taken complicating a direct measurement of the transient flux. We show the ATLAS and ZTF data together with the WISE flare for this source in the bottom right panel of Figure \ref{fig:optcounterparts}, which resemble those of typical optically-selected TDEs. Spectroscopically, the broad H$\alpha$ line in the classification spectrum did not show a strong blue absorption wing and P-Cygni profile that is commonly seen in spectra of Type II SNe \citep{Gutierrez2017}. Our 2023 spectrum also showed the formation of [Fe VII] coronal lines five years after the transient discovery, which has been seen in some TDEs in gas-rich environments \citep[e.g.][]{Onori2022,Short2023}, but is rare in Type II SNe \citep[e.g. SN2005ip;][]{Smith2009}. Another argument for the TDE nature of WTP18aampwj is that the total radiated energy in the IR alone exceeds $10^{51}$ erg already and is still ongoing. Likewise, the peak W1 luminosity exceeds what is seen in typical SNe \citep{Thevenot2020}. Thus, we suggest that this source is indeed a TDE rather than a supernova, as was also suggested in recent works based on the integrated IR luminosity \citep{Thevenot2020,Thevenot2021}. As the optical spectrum shows no evidence for AGN-like ionization levels, we place this source in our gold sample.

\end{appendix}

\bibliographystyle{aasjournal}
\bibliography{refs}

\end{document}